\documentclass[preprint]{elsarticle}

\usepackage{lineno,hyperref}
\modulolinenumbers[5]
\usepackage{xcolor}
\usepackage{booktabs}
\usepackage{subcaption}
\usepackage[section]{placeins}
\usepackage{rotating}
\modulolinenumbers[5]
\usepackage{ulem}
\journal{ }

\begin{document}

\begin{frontmatter}

\title{Identifying urban air pollution hot-spots by dispersion modeling when data are scarce: application to diesel generators in Beirut, Lebanon}

\author{Elissar Al Aawar$^1$, Abdelkader Baayoun$^1$, Alaa Imad$^1$, Jad El Helou$^1$, Lama Halabi$^1$, Mohamad Ghadban$^1$, Ali Moukhadder$^2$, Marya El Malki$^2$, Sara Najem$^3$, Najat A. Saliba$^4$, Alan Shihadeh$^1$}
\author[mymainaddress]{Issam Lakkis\corref{mycorrespondingauthor}}
\cortext[mycorrespondingauthor]{Corresponding author}
\ead{issam.lakkis@aub.edu}

\address{$^1$Department of Mechanical Engineering, Maroun Semaan Faculty of Engineering and Architecture, American University of Beirut, Beirut,1107 2020, Lebanon\\ $^2$ Department of Chemical Engineering, Maroun Semaan Faculty of Engineering and Architecture, American University of Beirut, Beirut, 1107 2020, Lebanon\\$^3$ National Center for Remote Sensing, National Council for Scientific Research, Riad al Soloh, 1107 2260, Beirut, Lebanon\\ $^4$ Department of Chemistry, Faculty of Arts and Sciences, American University of Beirut, Beirut, 1107 2020, Lebanon}

\begin{abstract}
	
Diesel generators are emerging  as community-initiated solutions to compensate for electricity shortage in cities marred by economical crisis and/or conflict. The resulting pollution distribution in dense urban environments is a major source of concern to the population. In the absence of periodic observations from properly distributed sensors, as is the case in Beirut, physically based computational modeling stand out as an effective tool for predicting the pollutant distribution in complex environments, and a cost-effective framework for investigating what-if scenarios and assessing mitigation strategies. Here, we present a Lagrangian transport model-based study of PM2.5 dispersion originating from a large number of diesel generators in Beirut. We explore large and small scale dispersion patterns in selected smalls domains and over the entire city. The scenarios considered investigate the impact of topography, atmospheric stability, presence of buildings, diesel generators distribution, and stacks elevations for representative meteorological conditions. Assessment of these scenarios is carried out in terms of small and large scale dispersion patterns and the mean concentration at street level and population exposure proxy indicators. We also report on the efficacy of elevating the stack height as a mitigation measure at different representative wind and atmospheric stability conditions.

\end{abstract}

\end{frontmatter}


\section{Introduction}\label{SecIntroduction}
Urban areas with high population density are linked with hotspots of air pollution \cite{du2019does,xian2019recent}. In fact, more than 80\% of people living in urban areas are exposed to air quality levels that exceed the World Health Organization (WHO) limits \cite{world2016ambient}. Fine particulate matter (PM$_{2.5}$) is one of the most critical air pollutants affecting human health \cite{brunekreef2002air,kappos2004health,schwartz1996daily}; therefore, continuous monitoring of PM$_{2.5}$ concentrations has attracted extensive attention for epidemiological studies in urban areas \cite{atkinson2014epidemiological, dockery1993association, fann2012estimating, wilson1997fine}. However, many developing cities cannot afford to install and maintain an expensive air quality monitoring network, and as an alternative, they resort to air quality modeling to predict PM$_{2.5}$ concentrations \cite{abdallah2018first}. The lack of efficient urban planning is associated with the abundance of street canyons in most cities. Street canyons tend to trap air pollutants and consequently deteriorate air quality, especially when the buildings are tall and the streets are narrow \cite{liu2004large}.To capture these effects on the pollution hot spots, the dispersion model must resolve the buildings that form the street canyons (at a scale $\sim 1 - 5$ m). The model must also take into account the prevailing wind conditions, the spatial distribution of the emission sources, their emission rates, in addition to the topography. Despite its limitations, the Gaussian plume model performs well in open terrains \cite{heist2013estimating} and satisfactorily captures the effect of atmospheric stability classes on plume dispersion \cite{abdel2008atmospheric}. On the other hand, contrary to the Gaussian plume model, which is not suitable for urban environments involving various street canyon configurations \cite{di2008flow}, Lagrangian particle dispersion models are driven by three dimensional building-resolving flow simulations which allow for a physically more accurate representation of pollution transport in these complex domains \cite{pullen2005comparison}.\\
The modeling system GRAMM (Graz mesoscale model (Oettl, 2017))-GRAL (Graz Lagrangian model (Oettl, 2018)) has been used to simulate the levels of various pollutants originating from different sources over entire cities in Europe. It was used for evaluation at the urban and meso scales in Zurich, Laussane (Switzerland), Vienna and Graz (Austria) and Po Valley (Italy) \cite{berchet2017cost, berchet2017evaluation, almbauer2000simulation, almbauer2000analysis, fabbi2019impact, kurz2014projection}. Other studies used GRAMM-GRAL to simulate pollution at specific road tunnel portals, highways on complex terrains, specific streets intersections and climate change mitigation in urban mobility~\cite{oettl2003dispersion, oettl2002simple, sturmpollutant, wolkinger2018evaluating}. The performance of this modeling system was evaluated using several settings of meteorological data, emission patterns and urban features at two levels. The first level deals with general model properties including convergence, grid characteristics and computational features. The second one deals with test cases of modeled data against real time measurements of wind data and pollutants concentrations. Both levels aim to prove the compliance of this model with the standard quality objectives and performance criteria. This was accomplished by different studies in the European countries and the model was found to be in compliance with the German Guideline VDI 3783 (2015) \cite{landesregierung2017documentation, landesregierung2018documentation} and to be fulfilling the criteria as defined by the Forum for Air quality Modelling in Europe (FAIRMODE) \cite{berchet2017evaluation}. Another study, conducted in Australia, subjected GRAMM-GRAL to a model-to-data comparison as well as a model-to-model comparison. Both models were compared to commonly used models in Australia; GRAMM was compared to CALMET and GRAL was compared to CAL3QHCR. GRAMM-GRAL was found to give results that are as good as those predicted by the other models \cite{optimization2017, optimization2017app}.   \\

Beirut, the capital of Lebanon, lying on the eastern shore of the Mediterranean Sea, is characterized by high PM$_{2.5}$ concentration levels \cite{massoud2011intraurban, saliba2010origin} exceeding the WHO limits by 150\% \cite{farah2018analysis}. These high concentrations are linked to long-range pollution transport such as the dust events originating in the Sahara and Arabian deserts \cite{lovett2018oxidative} and local-range pollution transport due to emissions from traffic \cite{daher2013chemical, massoud2011intraurban}, construction sites \cite{saliba2010origin}, and diesel generators \cite{daher2013chemical}.\\
Diesel generators are widely spread inside Beirut because of the daily power outage periods. They are among the major sources of anthropogenic heat, and their contribution to the heat island effect in Beirut city was shown to dominate other factors \cite{mg2020novel}. In addition, Diesel generators have been shown to increase the exposure to carcinogens in Beirut city \cite{daher2013chemical}. A survey conducted in 2017 \cite{baayoun2019emission} on the diesel generators installed in the Hamra neighborhood of Beirut showed that 53\% of the buildings were equipped with at least one diesel generator. Additionally, the survey showed that around 469 generators exist in the Hamra area (0.55 $km^2$). The daily fuel consumption of diesel generators in the whole city was found to be 747 metric tons on average. This estimate was based on the survey and the imported fuel data provided by the ministry of energy and water (MOEW) and the ministry of environment (MOE). In addition to Beirut, many urban environments are increasingly relying on diesel generators as back up energy sources to compensate for the electricity outages. These outages occur as a result of insufficient generation that does not meet the population's demand or as a result of a weak infrastructure for proper transmission and distribution or even rapid industrialization. It is worth noting that several countries in Africa and Asia suffer from both problems \cite{alam2017tracking}. For instance, countries like Syria and Iraq are now relying more on diesel generators to support their grid as a consequence of the recent or ongoing wars \cite{jassimenvironmental,newsdeeply2014}.\\
Up to our knowledge, very few studies have documented PM emissions or concentration levels from diesel generators in urban environments. One study (Centre for Science and Environment, 2018) conducted in Gurugram, India, showed that diesel generators contribute to about 30-40\% increase in PM$_{2.5}$ levels after three to four hours of operation. Another study done in Nigeria \cite{oguntoke2017degradation} quantified concentrations of different pollutants including particulate matter at various distances from diesel generators. These studies, however, resort to measurements that are spatially limited, and there are no reported studies in which the dispersion of PM$_{2.5}$ emitted from diesel generators is modeled at the pedestrian level (few meters above ground) over a whole city. \\
Using model-based predictions of PM$_{2.5}$ concentration levels originating from the diesel generators installed in Beirut, the objectives of this study are to 
(i) estimate the spatial distribution of the population density weighted  PM$_{2.5}$ concentration as a proxy for human exposure, 
(ii) gain physical understanding of the impact of and the interplay between the key predictor variables (wind speed and direction, atmospheric stability, and stack height) on the observed large and small scale patterns, and (iii) assess the efficacy of increasing the stack heights as a potential mitigation measure. \\
This paper is organized as follows. In section \ref{SecMethodology}, we present the problem setup, the methodology, and the input data. In the results section, we discuss the effect of different parameters on the observed patterns of the spatial distribution of the PM$_{2.5}$ concentration and on the total population weighted concentration at different elevations, taken as a proxy for exposure of the inhabitants. We conclude the results section by a qualitative comparison between the time averaged model-predicted PM$_{2.5}$ map and a previously reported NO$_2$ observations-based map~\cite{badaro2014geostatistical}. In section \ref{SecConc}, we present a summary of the conclusions and  propose future work.

\section{Methodology}\label{SecMethodology}
\begin{figure}[!htbp]
	\begin{center}
		\includegraphics[width=\columnwidth]{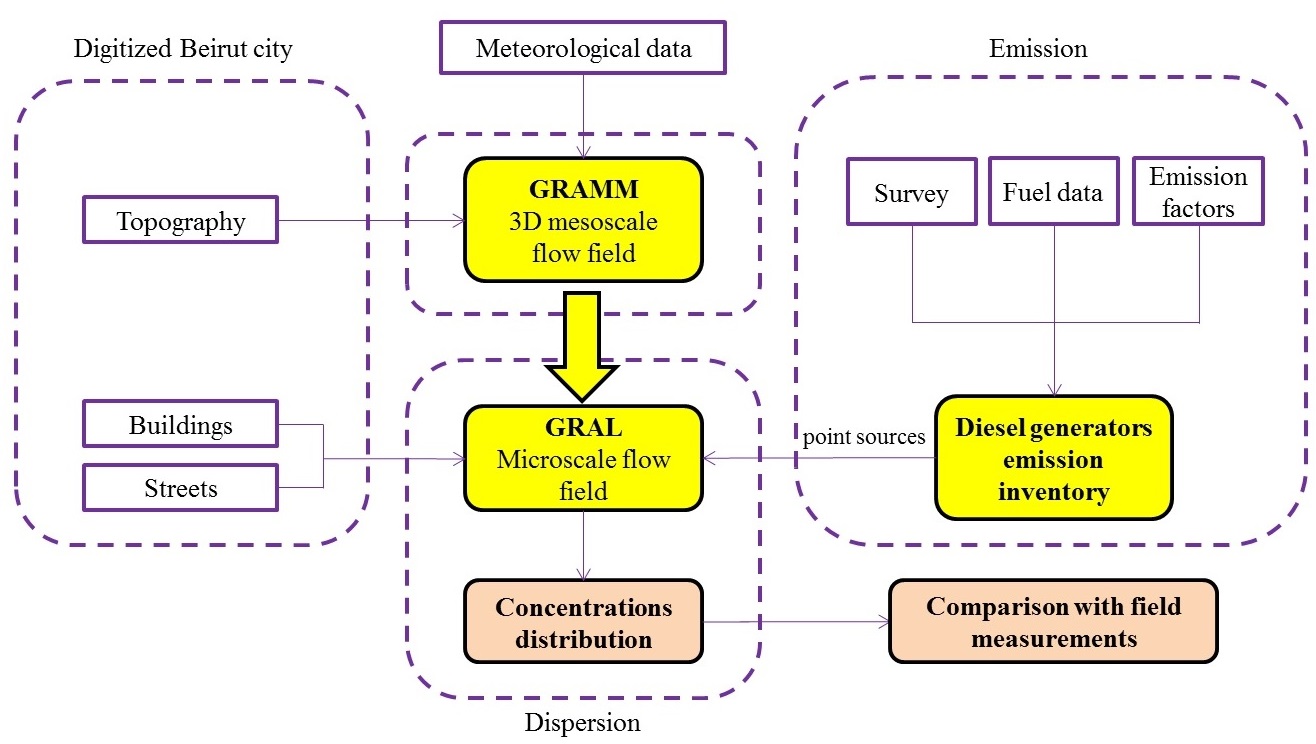}
		\caption{Methodology scheme of the applied model and input data}
		\label{FigMethodology}
	\end{center}
\end{figure}

The steady state pollutant concentration distributions are predicted using the GRAMM-GRAL tool (Version 18.01) at a high spatial resolution (2 m and 5 m) for representative monthly meteorological conditions, taking into account the topography, buildings geometries and distribution, atmospheric stability, distribution of diesel generators, and the stack height. For the scenarios where the buildings are included, the prognostic approach in the tool, where the flow is explicitly computed by forward integration of a set of prognostic equations, was employed. Figure \ref{FigMethodology} describes the methodology followed in this study to predict the spatial distribution of the PM$_{2.5}$ concentration at steady state representative of monthly meteorological conditions. First, information on Beirut City's topography, buildings, and streets were all digitized, and the emission inventory for diesel generators was built. The GRAMM module, which is a non-hydrostatic model, computes the mesoscale air flow-field in a larger domain (domain B) containing Beirut city, the domain of interest for species transport (domain A), as seen in Figure \ref{FigBeirutGG}-b. GRAMM takes as its input, in addition to the topography, wind speed (WSPD), wind direction (WD), and atmospheric stability class. Forced by GRAMM outputs, the GRAL model infers the microscale air flow-field inside the city while accounting for the effects of buildings on the flow and turbulence patterns. This microscale velocity field, along with the emission sources information, are then used by the Lagrangian dispersion module of GRAL to compute the concentration map in domain A.

\subsection{Digitizing Beirut City}\label{SubsecDigitization}
\begin{figure}[!htbp]
	\begin{center}
		\includegraphics[width=\columnwidth]{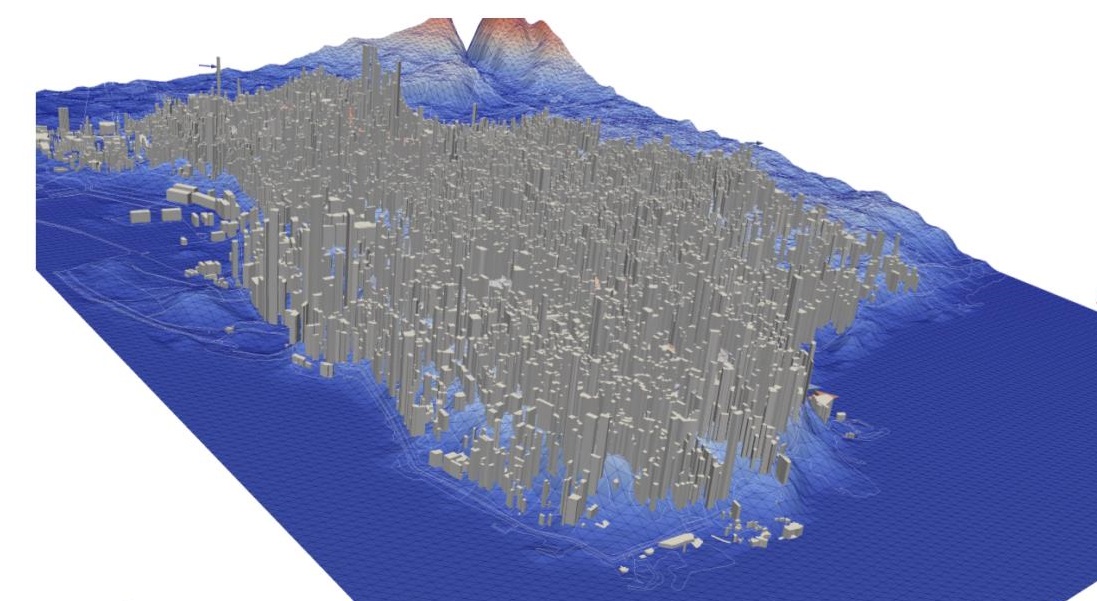}
		\caption{Digital representation (not to scale) of the topography, buildings, and streets of Beirut}
		\label{FigDigital}
	\end{center}
\end{figure}

Figure \ref{FigDigital} shows a digital representation of the topography, buildings, and streets of Beirut. The topography data, which was extracted at 50 m horizontal resolution for the Greater Beirut area \cite{gpsv}, was used as an input to the GRAMM domain. Overlaid are 18478 buildings, provided as a shapefile by the Lebanese army, and the streets of Beirut downloaded as OSM files from OpenStreetMap, and then converted to a shapefile using Quantum Geographic Information System (QGIS). Note that the the average building height and the standard deviation are 19 m and 12.88 m respectively.

\subsection{Domain Setup for Pollution Transport Simulations}\label{SubsecDomain}
\begin{figure}[!htbp]
	\begin{center}
		\includegraphics[width=\columnwidth]{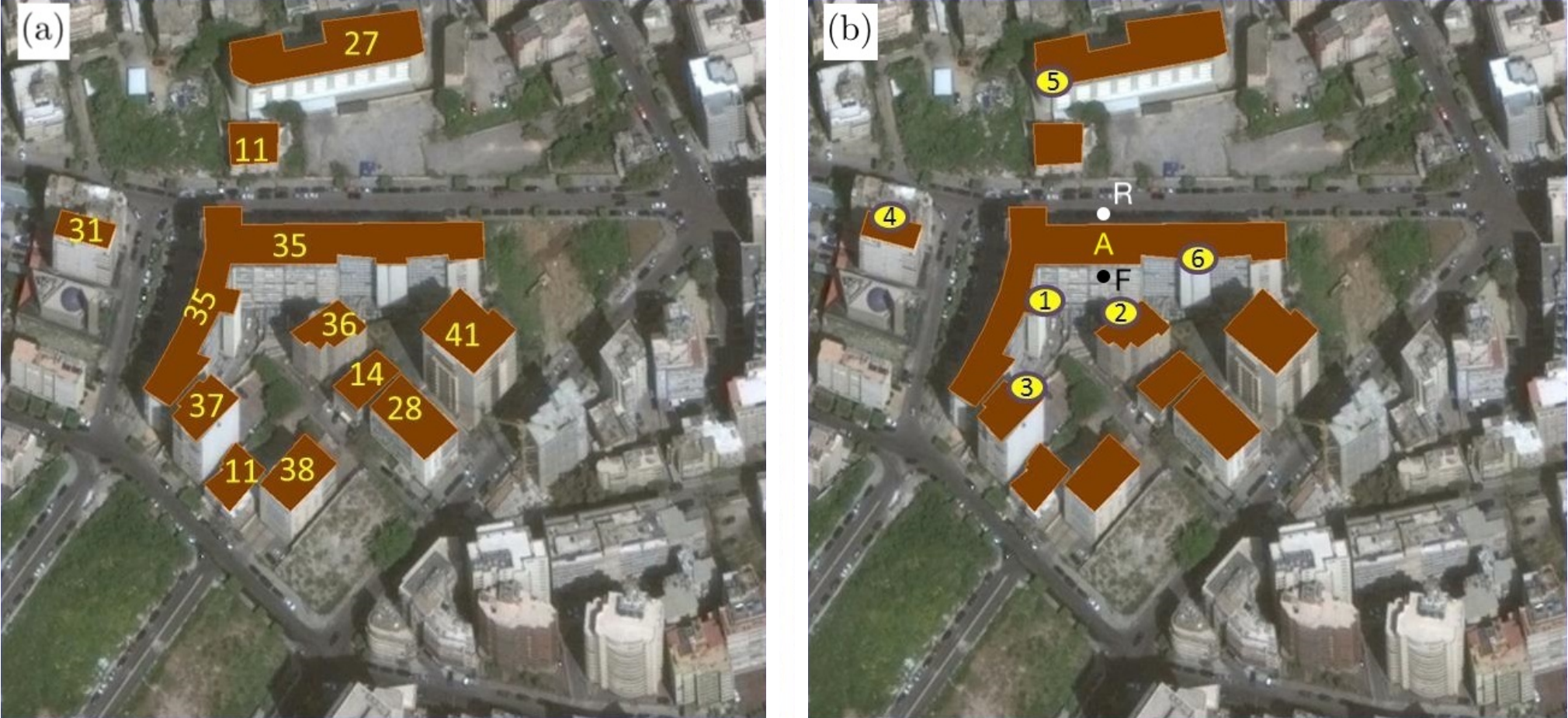}
		\caption{GRAL domain containing buildings in the Karakas neighborhood of Beirut showing (a) the height of each building labeled in meters and (b) the distribution of the diesel generators in the area.}
		\label{FigKarakas}
	\end{center}
\end{figure}

\begin{figure}[!htbp]
	\begin{center}
		\includegraphics[width=\columnwidth]{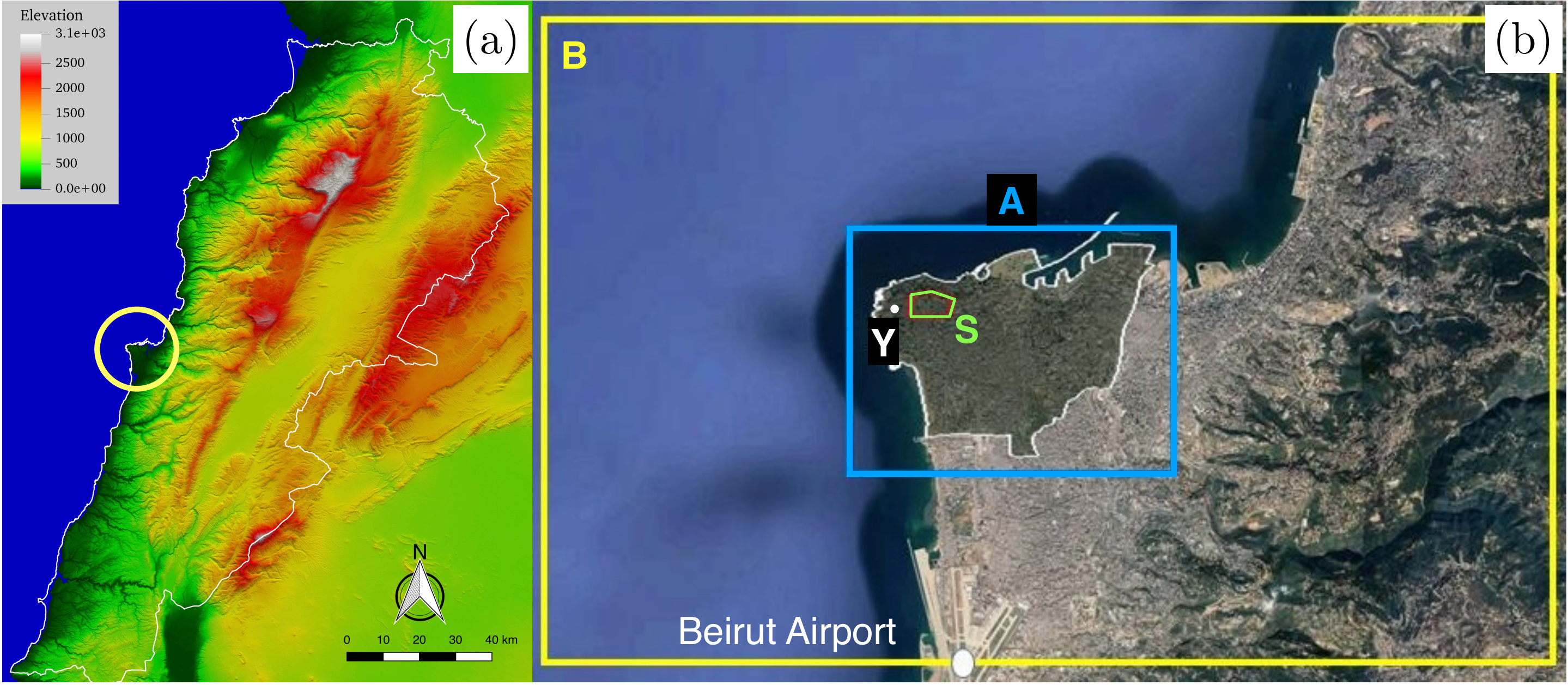}
		\caption{(a) Color map topography of Lebanon. (b) GRAMM (yellow) and GRAL (blue) domains for Beirut city. Border for Beirut is delimited by the white line while the border for the Hamra neighborhood is delimited in green.}
		\label{FigBeirutGG}
	\end{center}
\end{figure}

For the purpose of investigating the factors that determine the characteristics of pollutant distribution and identifying the governing patterns, we chose to study the transport in two domains. The first domain is a small domain that contains few buildings whereas the second is a large domain that contains the whole city of Beirut. In both cases, we study the transport of pollutants emitted from the diesel generators. As discussed later, the observations we make upon investigating the transport in the small domain and the conclusions we draw are instrumental in explaining the observed behavior and identifying the characterizing patterns of the pollutant distribution over the entire city, predicted for the various scenarios listed in Table \ref{table:CasesLarge}.  These two study domains are described next.\\
The small domain, located in the Karakas neighborhood of Beirut,  is shown in Figure \ref{FigKarakas} and  labeled Y in Figure \ref{FigBeirutGG}-b. In this domain. the dispersion of pollutants emitted by the generators is modeled at a 2 m resolution. The reason we chose this particular domain for the study is that the residents of the L-shaped 35 m high building reported high levels of particulate matter deposited on the fabric and furniture in their apartments. Also shown in Figure \ref{FigKarakas}-b are the locations of six diesel generators in the neighborhood, two of which are installed in the L-shaped building.\\
 
The fuel consumption, stack height, and the PM$_{2.5}$ emission rate of the diesel generators, represented as point sources in GRAL, are listed in Table \ref{table:Karakas}. Assuming that these generators work together for 3 hours per day, PM$_{2.5}$ emission rates were calculated according to the European Monitoring and Evaluation Programme/European Environmental Agency (EMEP/EEA) air pollutant emission inventory guidebook using the Tier 2 method for reciprocating engine applications \cite{european2016}.\\

\begin{table}[!htb]
	\centering
	\begin{tabular}{cccc}
		\toprule
		Label & Fuel consumption & Stack Height & PM$_{2.5}$  \\
		&  (Liters/month) &  (m) &  (kg/hour)  \\
		\midrule
		1 & 1800 & 37 & 0.0218 \\
		2 & 1000 & 38 & 0.0121 \\
		3 & 1500 & 39 & 0.0182 \\
		4 & 1000 & 33 & 0.0121 \\
		5 & 1000 & 29 & 0.0121 \\
		6 & 1800 & 37 & 0.0218
		\\
		\bottomrule
	\end{tabular}
	\caption{Data of the diesel generators distributed in the Karakas neighborhood.}
	\label{table:Karakas}
\end{table}

The large domain (domain A), used to study the dispersion of pollutants emitted by generators in the city of Beirut at a 5 m resolution, is bounded by the small blue rectangle in Figure \ref{FigBeirutGG}-b. As mentioned earlier, GRAMM requires a larger enclosing domain (domain B) to provide the velocity boundary conditions for domain A. The resolution of the GRAMM domain, bounded by the large yellow rectangle in Figure \ref{FigBeirutGG}-b, is set to 50 m. As can be seen in the topography color map in Figure \ref{FigBeirutGG}-a, Beirut is nearly at the center of a narrow coastal plain (210 km long and approximately 3 km wide) that stretches in a south-north direction. The city overlooks the Mediterranean Sea to the west and north. To the east, it is bounded by the western Mount Lebanon range which extends along nearly 170 km parallel to the Mediterranean coast, with peaks reaching $\sim3$ km.\\

Based on the survey on diesel generators in Hamra (red polygon labeled S in Figure \ref{FigBeirutGG}-b), the distribution of the generators in the rest of the city considers the capacities of the generators in a statistical approach embracing the probability of a generator in a specific location to have a specific capacity (in kVA). Then, this capacity is proportionally converted into fuel consumption while preserving the average daily fuel consumption of generators in Beirut (747 metric tons). The PM$_{2.5}$ emission rate is then estimated using an average value of 1.2999 kg of PM$_{2.5}$ emitted per one metric ton of fuel consumption\cite{european2016}.  To investigate the sensitivity of the PM$_{2.5}$ concentration spatial distribution to the generators locations, a second distribution of generators in Beirut was considered. Based on the survey in Hamra, this first order approximate distribution assumes one generator per two buildings in the rest of the city, making the number of diesel generators distributed in Beirut about 9369. As argued in section Appendix \ref{appA}, the assumption is justified, in part, due to the absence of availability of data on the various factors contributing to the generators spatial density. Since the exact location of these generators is not known outside Hamra, the generators were randomly placed at the centroids of the buildings footprints (GIS shapefile). The generators fuel consumption was also randomly distributed in a manner that conserves the total daily fuel consumption in Beirut. For both distributions,  all the diesel generators were assumed to operate for twelve hours per day. The two distributions, referred to as GD1 and GD2 respectively, are discussed in Appendix \ref{appA}. As for the locations of the stacks exhausts, the following two scenarios were considered: (a) the exhausts are located at the centroids of the buildings 2 m above the rooftop (RT+2) and (b) the exhausts are located at the street level 4 m above the ground (G+4). Scenarios 10-17 listed in Table \ref{table:CasesLarge} study the effect of buildings, atmospheric stability and generators distribution, whereas scenarios 17-19 study the effect of stack heights, and scenarios 20-29 study the effect of wind direction and speed.

\subsection{Meteorological conditions}\label{SubsecMeteo}
In addition to the emission sources, topography and buildings geometries, the dispersion model takes as input the wind speed, wind direction, and atmospheric stability conditions.  Noting that the predominant average wind speed is 4 m/s (at 10m anemometer level)\cite{wf}, wind speeds ranging from 1 to 4 m/s were considered  for the large domain. The following prevailing wind directions were considered: 247$^{\circ}$ (August and September), 225$^{\circ}$ (January - July and December), and 337$^{\circ}$ (October and November). With an annual average of 4m/s and 247$^{\circ}$, these values are monthly averages of measurements recorded by the weather station at Beirut Airport between June 2005 and September 2018, daily from 7am to 7pm \cite{wf}.  Unstable, and slightly unstable atmospheric stability classes were taken in the simulations. The choice of these specific stability classes is based on the Pasquill table \cite{berchet2017evaluation}. The different scenarios are summarized in Table \ref{table:CasesSmall} for the small domain and in Table \ref{table:CasesLarge} for the large scale domain. Note that in addition to 247$^{\circ}$, an additional wind direction of 270$^{\circ}$ was also considered for the small domain for sensitivity analysis. 

\begin{table}[!htb]
	\centering
	\begin{tabular}{cccc}
		\toprule
		Scenarios &  Wind Direction & Atmospheric stability & Increase in exhaust   \\
		&    ($^{\circ}$)  & class & elevation (m) \\
		\midrule
		1 &   247 & SU  & 0 \\
		2 &   247 & U & 0 \\
		3 &   247 & SU	& 0 \\
		\midrule
		4 &   247	& U	& 2 \\
		5 &	  247	& U	& 4 \\
		6 &  247	& U	& 6 \\
		7 &	 270	& U	& 2 \\
		8 &	  270	& U	& 4 \\
		9 & 	  270	& U	& 6 
		\\
		\bottomrule
	\end{tabular}
	\caption{Scenarios considered for the small domain. Wind speed is fixed at 4 m/s. U: unstable, SU: Slightly unstable. Scenarios 1-3 study the effect of atmospheric stability class. Scenarios 4-9 study the effect of wind direction and increasing the exhaust elevation.}
	\label{table:CasesSmall}
\end{table}

\begin{table}[!h]
	\centering
	\begin{tabular}{ccccccccc}
		\toprule
		Scenario & Wind & Wind & Atmospheric   & Distribution    & Exhaust   & Buildings  \\
		& Speed & Direction   & Stability &	of the  &  elevation    & Included?\\
		& (m/s) &  ($^{\circ}$)    &  Class     & generators & 	     (m)      &   \\
		\midrule
		10 & 4 & 247 & U & GD1   & RT+2 & N  \\
		11 & 4 & 247 & U & GD1   & RT+2 & Y  \\
		12 & 4 & 247 & SU	& GD1 & RT+2 &	N  \\
		13$^\star$ & 4 & 247 &SU	& GD1 &	RT+2 &	Y \\
		\midrule
		14 & 4 & 247 & U & GD2  & RT+2 & N  \\
		15 & 4 & 247 & U & GD2   & RT+2 & Y  \\
		16 & 4 & 247 & SU & GD2 & RT+2 & N \\
		17$^\star$ & 4 & 247 & SU & GD2   & RT+2 & Y  \\
		\midrule
		18	& 4 & 247 &	SU	& GD1 & G+4 &	Y \\
		19	& 4 & 247 &	SU	& GD2  &	G+4 &	Y \\
		\midrule
		20  & 4 & 225 & SU & GD1 & RT+2 & N \\
		21  & 4 & 225 & SU & GD1 & RT+2 & Y \\
		\midrule
		22  & 4 & 337 & SU& GD1 & RT+2 & N \\
		23  & 4 & 337 & SU & GD1 & RT+2 & Y\\
		\midrule
		24  & 3 & 247 & SU & GD1 & RT+2 & N \\
		25   & 3 & 247 & SU & GD1 & RT+2 & Y   \\
		\midrule
		26  & 2 & 247 & SU & GD1 & RT+2 & N  \\
		27  & 2 & 247 & SU & GD1 & RT+2 & Y \\ 
		\midrule
		28  & 1 & 247 & SU & GD1 & RT+2 & N  \\
		29  & 1 & 247 & SU & GD1 & RT+2 & Y  \\
		\midrule
		30  & 1 & 247 & SU & GD1 & G+4 & Y \\
		\midrule
		31	& 4 & 247 &	SU	& GD2 & G+4 &	Y \\
		\bottomrule
	\end{tabular}
	\caption{Scenarios investigated in the large domain. U: unstable, SU: slightly unstable, GD1: statistical distribution of generators locations and capacities, GD2: first order approximate distribution of generators locations and capacities, G:  ground level, RT roof top. $\star$ refers 
		to the base case. $\pi_1$: longitude-latitude integral of population weighted PM$_{2.5}$ concentration at 1.75 m above ground (mean elevation of the first floor), $\Pi$: sum of $\pi_k$ over all floors $k$.}
	\label{table:CasesLarge}
\end{table}

\subsection{Population Density Distribution}\label{SubsecPop}
As a proxy for human exposure to PM$_{2.5}$, we present spatial distributions of the PM$_{2.5}$ concentration weighted by the population density at different elevations corresponding to the different floors comprising the buildings. To this end, the concentration maps are extracted at elevations ranging from 1.75 meter up to 60 meters with an increment equal to the average floor height of 3.5 meters. Since the spatial distribution of the population density is not made available by the government\footnote{Another indication of the level of scarcity of data in cities like Beirut.},  we synthesized it from the electricity consumption data collected from Electricite du Liban (EDL). The number of inhabitants in a building was determined as the ratio of the monthly electricity consumption of the building to monthly consumption per capita, estimated to be 241 kWh according to the the World Bank collection of development indicators in Lebanon\cite{tradingeconomics}. The building population is then assumed to be uniformly distributed among the storeys.

\section{Results and Discussion}\label{SecResults}
In this section, we first explore the impact of wind direction and atmospheric stability on the spatial distribution of the PM$_{2.5}$ concentration predicted by the Lagrangian particle dispersion model (GRAMM-GRAL) in the small domain shown in Figure \ref{FigKarakas}. We also assess the efficacy of increasing the elevation of the emission sources as a mitigating measure to reduce the exposure of residents. The conclusions we draw from the discussion will, in part, help explain the characteristics of the dispersion patterns over the entire city for the scenarios listed in Table \ref{table:CasesLarge}. 

\subsection{Small Domain}\label{SubsecSmallDomain}
\begin{figure}[!htb]
	\begin{center}
		\includegraphics[width=\columnwidth]{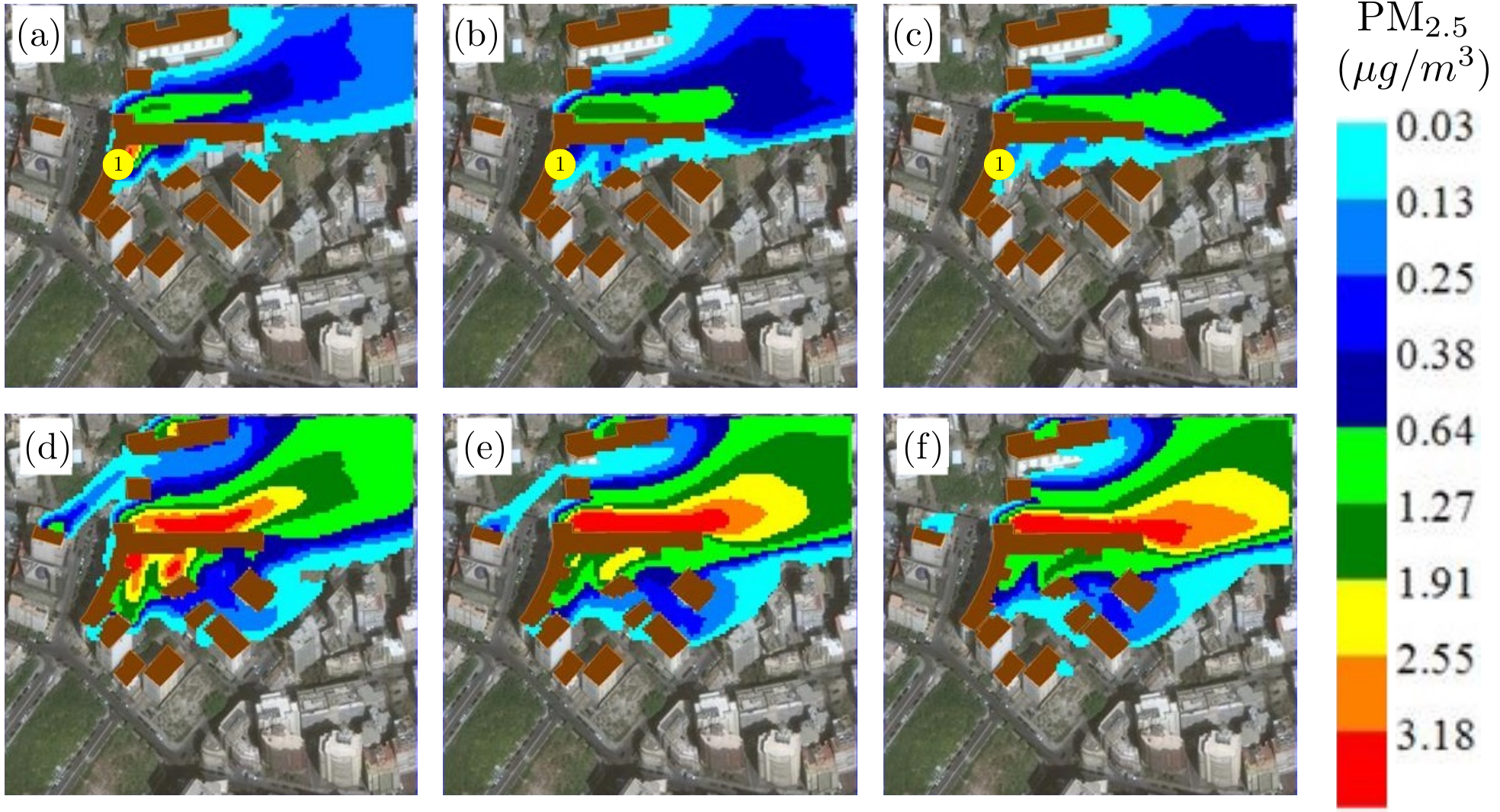}
		\caption{Top: Plume dispersion when only diesel generator \#1 is operating for (a) strongly unstable (b) unstable and (c) slightly unstable atmosphere. 
			Bottom: Plume dispersion when all generators are running together for (d) strongly unstable (e) unstable and (f) slightly unstable atmosphere. Concentrations of PM$_{2.5}$ are given at 5 m above ground. Wind direction (gray arrow) is 247$^\circ$. All the generators are running.}
		\label{FigKarakas1}
	\end{center}
\end{figure}

Figures \ref{FigKarakas1}a-c show the steady state dispersion of the plume when only diesel generator 1 (refer to Figure \ref{FigKarakas}-b) is operating. When the atmosphere is strongly unstable (Scenario  \#1, Table \ref{table:CasesSmall} ), a small local peak is observed near the stack. However, for a less unstable atmosphere (Scenarios  \#2 and 3), a large concentration zone of PM$_{2.5}$ is seen farther downward at the rear side of the building. This behavior is still observed when all the diesel generators are running together as presented in Figure \ref{FigKarakas1}d-f. \\
Due to strong mixing, turbulence in the unstable atmosphere increases the concentration levels in the immediate neighborhood of the stack.  Farther downwind, however, concentrations drop off very quickly.  In contrast, the mixing rate of the plume as it travels downwind is less when the atmosphere is stable.  This results in concentrations that are smaller in the neighborhood of the stack and higher farther downstream.\\

\begin{figure}[!htb]
	\begin{center}
		\includegraphics[width=\columnwidth]{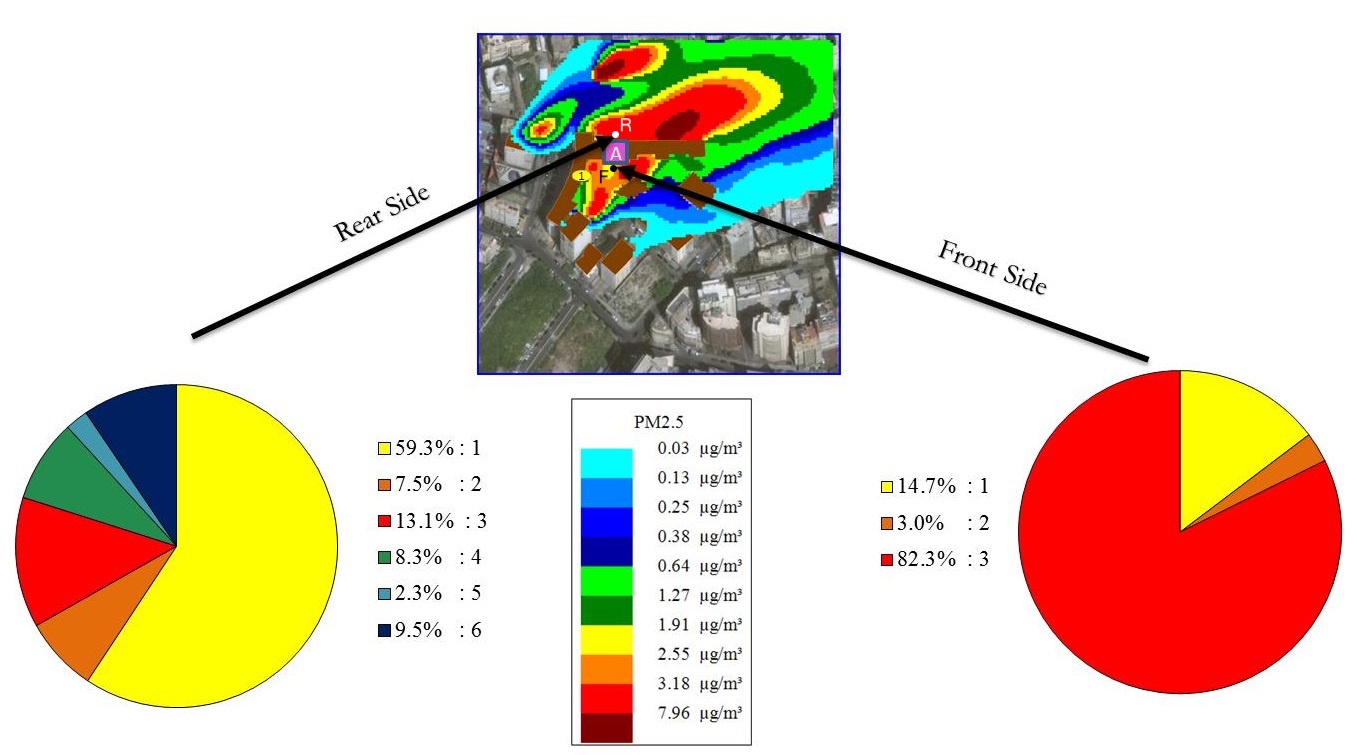}
		\caption{Source apportionment for the PM$_{2.5}$ concentration levels at the front side (location ``F'') and rear side (location ``R'') of an apartment ``A'' located at 30 m above ground. Unstable atmosphere. Wind direction is 247$^\circ$. All the generators are running.}
		\label{FigKarakas3}
	\end{center}
\end{figure}

Figure \ref{FigKarakas3} shows a source apportionment of the PM$_{2.5}$ concentration measured at the front side (location ``F'') and the rear side (location ``R'') of the apartment labeled ``A'' located at the 10th floor, i.e. 30 m above ground. It was found that generator \#1, facing the front side of the apartment along the southwest direction, contributes to nearly 60\% of the predicted concentration at the rear side. Its contribution to the front side, however, is only 15\%.  On the other hand, Generator \#3, which also faces the front side along the southwest direction but is 30 m farther than Generator \#1, contributes to nearly 82\% of the concentration level measured at the front side and to only 13\% at the back side. These observations reveal the complexity of pollution transport in urban environments. Although both generators have the same stack height (elevation of the emission source) and are on the southwest of the front side of apartment ``A'', the closer generator contributes most to the backside (``R'') while the farther generator contributes most to the front side (``F''). This behavior is attributed to the downwash effect of the plume. In a given horizontal plane below the emission source, the location of the highest concentration zone is largely dictated by the velocity field, which in turn depends on the incoming wind speed and direction, the stability of the boundary layer, and the buildings geometries and orientations.\\

\begin{figure}[!htb]
	\begin{center}
		\includegraphics[width=\columnwidth]{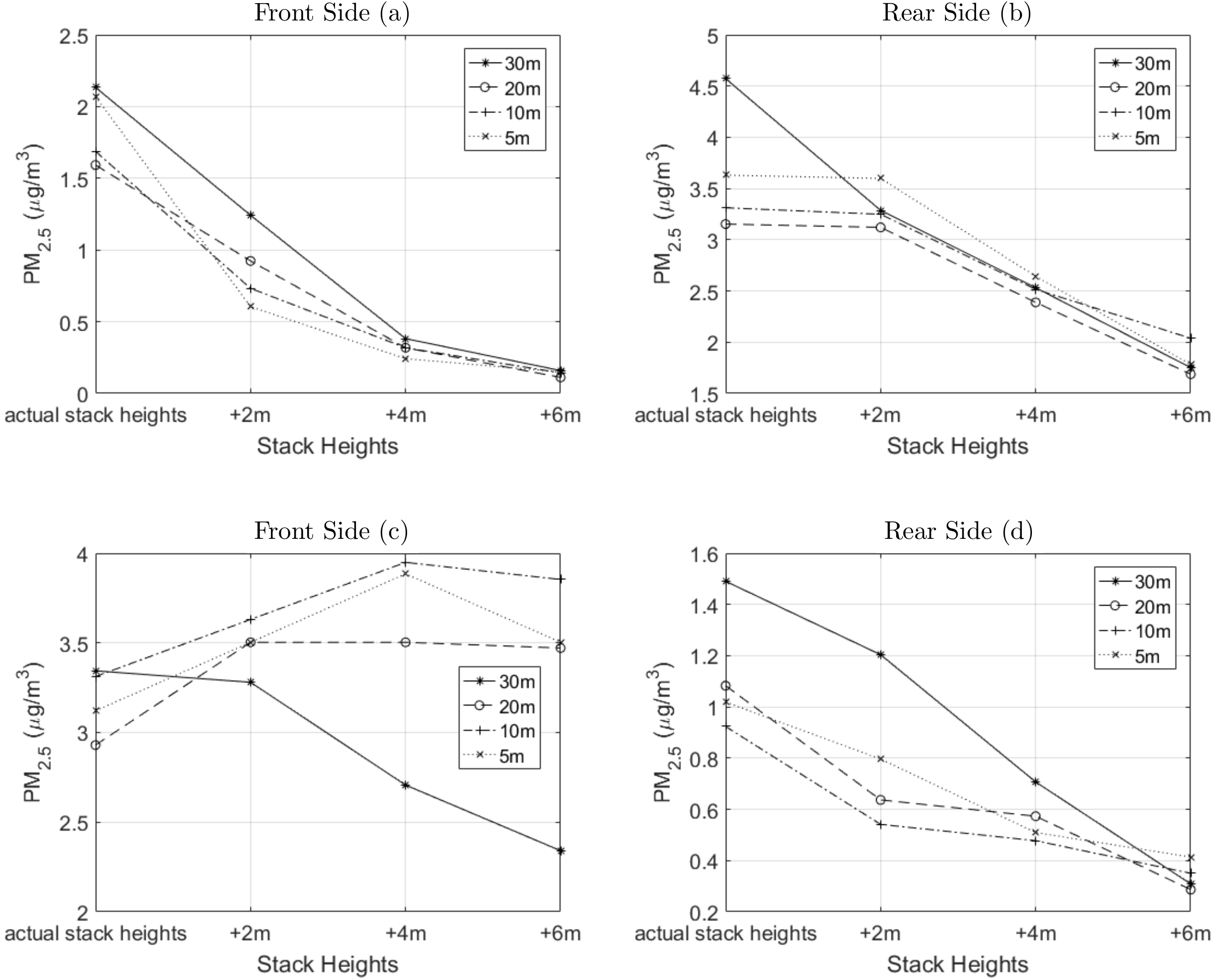}
		\caption{Effect of stack height increase on PM$_{2.5}$ concentrations at the front and rear sides of apartment ``A'' and lower levels for 247$^\circ$  (a-b) and 270$^\circ$ (c-d) wind directions.}
		\label{FigKarakas4}
	\end{center}
\end{figure}

In the absence of buildings and when the topography is flat, increasing the stack height is expected to result in lower concentration levels in the far field at elevations below the stack, which is in accordance with the Gaussian plume model \cite{abdel2008atmospheric}. Here, we explore whether this observation remains true for multiple sources in the presence of buildings. To this end, we assess the impact of increasing the stack height on the PM$_{2.5}$ concentration levels at different elevations on the vertical lines passing through points ``F'' and ``R'' (see Figure 3-b). Figures \ref{FigKarakas4}a-b shows that, for a wind direction of 247$^\circ$, extending the stack height (up to 6 m) results in considerable reduction in the PM$_{2.5}$ concentration levels at all elevations considered. On the other hand, for a wind direction of 270$^\circ$, increasing the stack height increases the PM$_{2.5}$ concentration levels at elevations below 30 m on the front side, as shown in Figures \ref{FigKarakas4}c-d. Therefore, in the presence of buildings and multiple emission sources, moderately increasing the stack height does not necessarily reduce the concentration levels at all elevations in the neighborhood. \\

The pollutant concentration field in the urban environment is a function of the complex interplay between source location and emission rate, atmospheric stability, buildings geometry and orientation, and wind direction and speed. This renders simple mitigating measures, such as increasing stack height or changing its location, of limited effectiveness.

\subsection{Large Domain}\label{SubsecLargeDomain}
\begin{figure}[!htb]
	\begin{center}
		\includegraphics[width=3.4in]{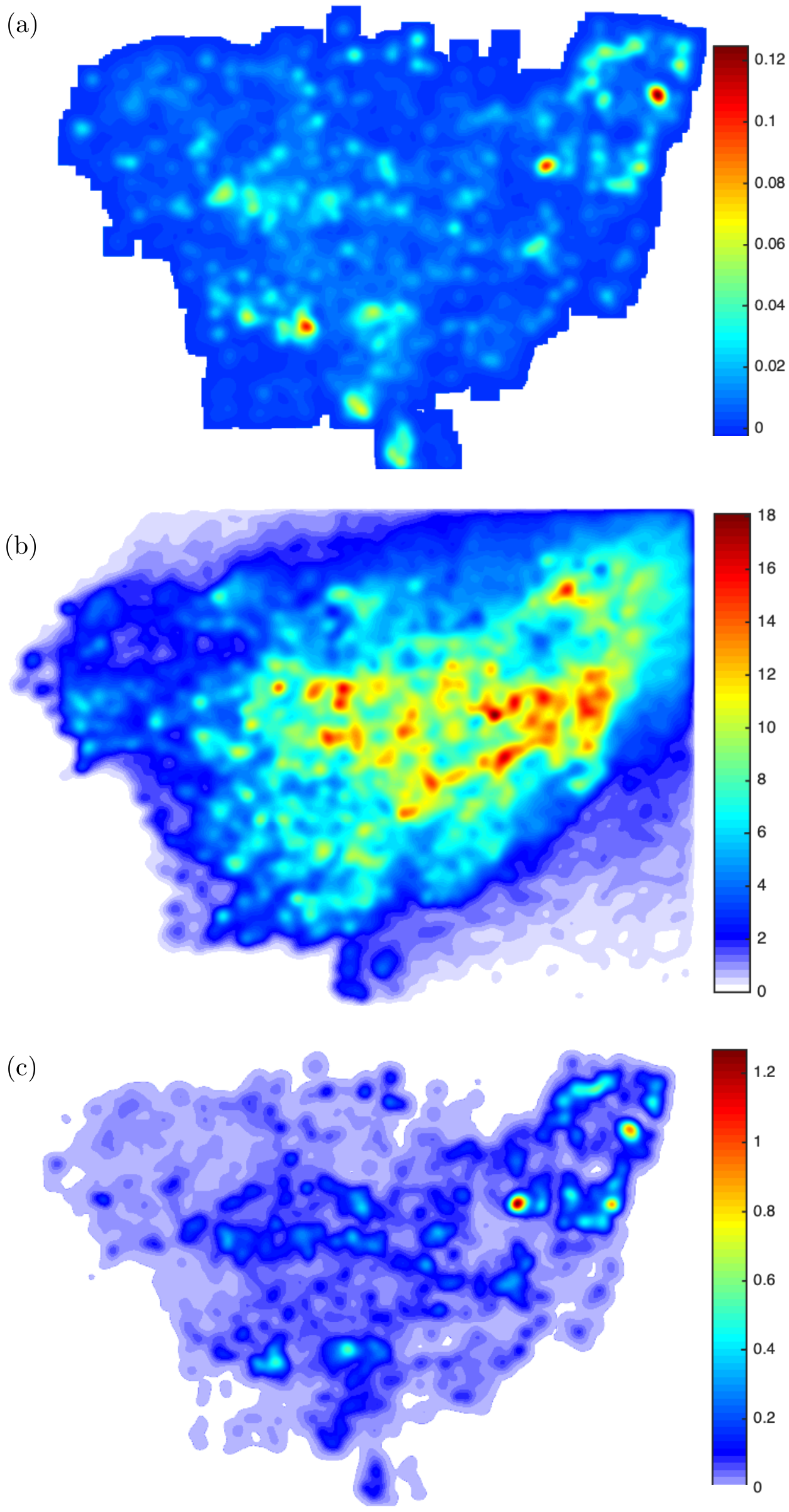}
	\end{center}	
	\caption{Spatial distributions at $z_1=$1.75 m elevation (floor 1) over the longitude-latitude plane of Beirut of (a) the smoothed population density ($\tilde{\rho}(x,y,z_1)$),  (b) smoothed PM$_{2.5}$ concentration ($\tilde{PM}_{2.5}(x,y,z_1)$), and (c) their product ($\tilde{PM}_{2.5}(x,y,z_1) \tilde{\rho}(x,y,z_1)$) for the base case Scenario 13.}
	\label{pops}	
\end{figure}

In this section, we investigate the impact of (i) topography, (ii) atmospheric stability, (iii) presence of buildings, (iv) diesel generators distribution, (v) emission sources elevations (stacks heights), and (vi) wind conditions on the concentration field in the large domain. The discussion involves exploring the characteristics of the persistent hot pollution spots and those of the persistent cold spots. In addition, we quantify the impact of the various conditions by comparing the corresponding scenarios in terms of the population-weighted PM$_{2.5}$ concentration and the mean concentration at street level.\\
The mean concentration at the elevation $z$ above the ground is computed as 
\begin{eqnarray}
\overline{PM}_{2.5}(z) = \frac{\int_{\Omega_R} PM_{2.5}(x,y,z) dx dy }{\int_{\Omega_R} dx dy}
\end{eqnarray}
where $\Omega_R$ is the longitude-latitude domain that constitutes of all non-built up space that is within a distance $R$\
from the nearest building. For mean concentration at street level, we take $z= 2m$ and $R=25 m$.\\
The  spatial integral of the PM$_{2.5}$ concentration weighted by the population density $\rho$, expressed as  
\begin{eqnarray}
\pi_k = \int \int \tilde{PM}_{2.5}(x,y,z_k) \tilde{\rho}(x,y,z_k) \, dx \, dy,
\end{eqnarray}
is considered as a proxy for human exposure at different elevations $z_k$ (corresponding to average elevations of the building storeys, $k=1..N_S$). To ensure spatial overlap between the two quantities (available on a grid as output from the dispersion model), the PM$_{2.5}$ concentration and the population density, $\rho$,  are smoothed according to
\begin{eqnarray}
\tilde{PM}_{2.5}(x,y,z_k) & = & \sum_{i} \sum_{j}  {PM}_{2.5}(x_i,y_j,z_k) \Delta x_g \Delta y_g \, \phi_{\sigma}(x-x_i,y-y_i) \\
\tilde{\rho}(x,y,z_k) & = & \sum_{i} \sum_{j}  \rho(x_i,y_j,z_k) \Delta x_g \Delta y_g \, \phi_{\sigma}(x-x_i,y-y_i)
\end{eqnarray}
where $(x_i,y_j)$ is the position of grid cell $(i,j)$, $\Delta x_g$ and $\Delta y_g$ are the dimensions of the grid cell, and  $\phi_{\sigma}$ is a smoothing (Gaussian) function characterized by smoothing radius $\sigma$
\begin{eqnarray}
\phi_{\sigma}(x,y) = \frac{1}{\pi \sigma^2} \rm{exp}^{-\frac{x^2+y^2}{\sigma^2}}.
\end{eqnarray}
Note that $\pi_k$ and its sum over all building storeys $\Pi = \sum_{k=1}^{N_S} \pi_k$, normalized by their values for the base cases (scenario 13 for GD1 and scenario 17 for GD2), are reported in the last two columns of table \ref{table:CasesLarge}. For the base case (scenario 13), the spatial distributions (over the longitude-latitude plane of Beirut) at 1.75 m elevation (floor 1) of the smoothed population density,  smoothed PM$_{2.5}$ concentration, and their product are shown in Figs \ref{pops}-a, b, and c respectively. \\
We conclude this section by a qualitative comparison of model predictions with the NO$_2$ map of Beirut reported in ~\cite{badaro2014geostatistical}.

\subsubsection{Effect of Topography}
\begin{figure}[!htb]
	\begin{center}
		\includegraphics[width=\columnwidth]{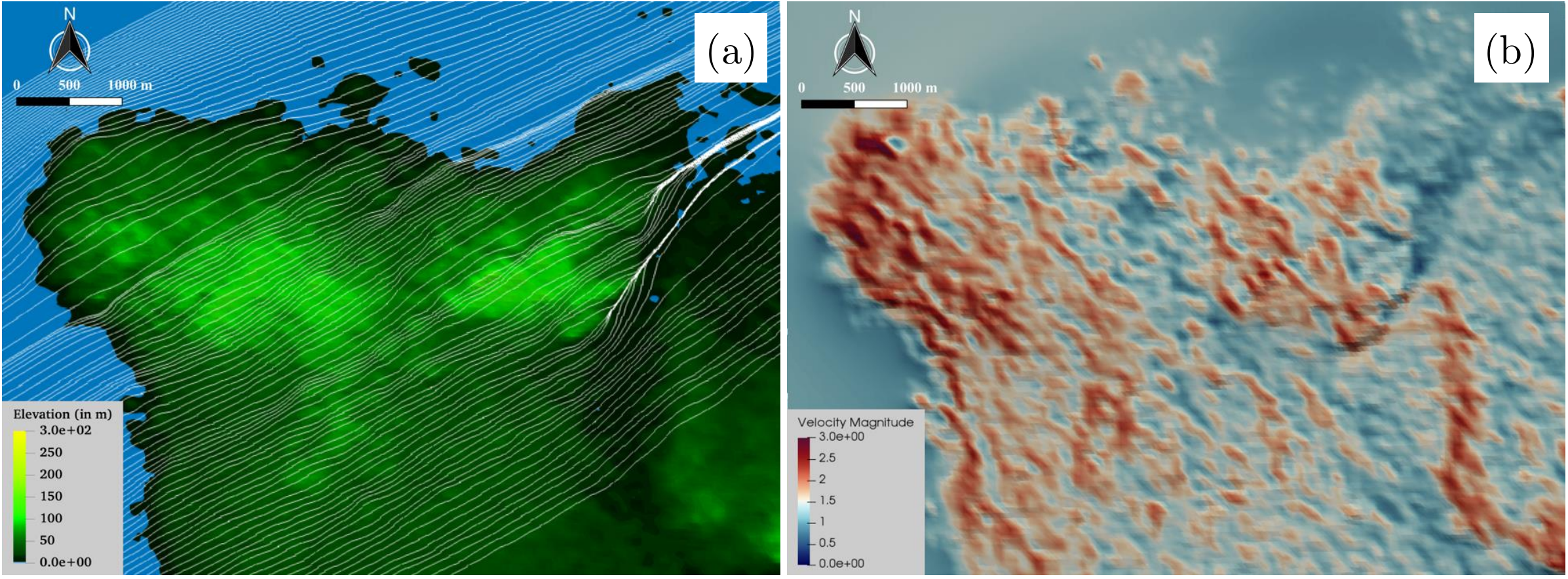}
		\caption{(a) Topography and streamlines visualization of the steady-state wind velocity field at 2 m above ground in the GRAL domain, (b) Magnitude of the steady- wind velocity field at 2 m above ground. Slightly unstable atmosphere.}
		\label{FigBeirutTopStream}
	\end{center}
\end{figure}

The topography plays a key role in the large-scale pollution transport. This is because the pollutant is advected by the velocity field, which in turn depends on the topography. Based on the law of mass conservation, one expects high wind speeds on the hills and low wind speeds in the valleys. In the absence of obstacles, the pollutant residence time is largely decided by the wind speed so that the pollutant gets swept away quickly from regions of high wind speed and tends to accumulate in regions where the wind speed is low. Bounded by the Mediterranean sea to the west and by Mount Lebanon to the east, the topography of Beirut is characterized by two hills, each at approximately 100 m above sea level, separated by a low-elevation pass as depicted in Figure \ref{FigBeirutTopStream}-a. \\
The resulting velocity field, shown in Figure \ref{FigBeirutTopStream}-b, shows that lower speeds are observed in the east side of the city, which is characterized by a downward slope created by the river of Beirut that separates the district of Matn in the Mount Lebanon Governorate from Beirut. Higher wind speeds are observed in the western side, especially in at the northwest near the sea. By considering only the incoming wind conditions and the topography, we expect higher pollution levels in the low wind speed regions of Figure \ref{FigBeirutTopStream}-b.

\subsubsection{Effect of Buildings}
The spatial distributions of the PM$_{2.5}$ concentration at 2 m above the ground for the base case (scenario 13) is shown in Fig. \ref{FigPM2P5B1}-b. For this scenario, the buildings are included, the wind speed and direction are 4 m/s and 247 degree respectively, the atmospheric stability class is ``slightly unstable", and the generators distribution is GD1 with stacks located at 2 m above rooftops. The first floor and total exposure proxy indicators ($\pi_1$ and $\Pi$) for all the scenarios that employ the GD1 generators distribution are normalized by their values ($\pi_1^*$ and $\Pi^*$) for this base case. In the absence of buildings, while keeping all the other conditions the same (scenario 12), the PM$_{2.5}$ concentration map is shown in Fig. \ref{FigPM2P5B1}-a. When comparing the two distributions, one can observe that the presence of the buildings results in a highly non-uniform distribution characterized by many islands of elevated PM$_{2.5}$ concentration.  The presence of the buildings is also responsible for  28\% and 15\% rises in the  first floor and total exposure proxy indicators respectively; as can be inferred from the lower values of the normalized first floor and total exposure proxy indicators $\pi_1/\pi_1^*=0.78$ and $\Pi/\Pi^*=0.87$ in scenario 12. In addition, the presence of the buildings resulted in around 25\% increase in $\overline{PM}_{2.5}^{(25 m)}(2m)$ the average PM$_{2.5}$ concentration, which is close to 25\%, the reduction in non-built up area brought about by adding the buildings. 

\subsubsection{Effect of Atmospheric Stability Class}
\begin{figure}[!htb]
	\begin{center}
		\includegraphics[width=\columnwidth]{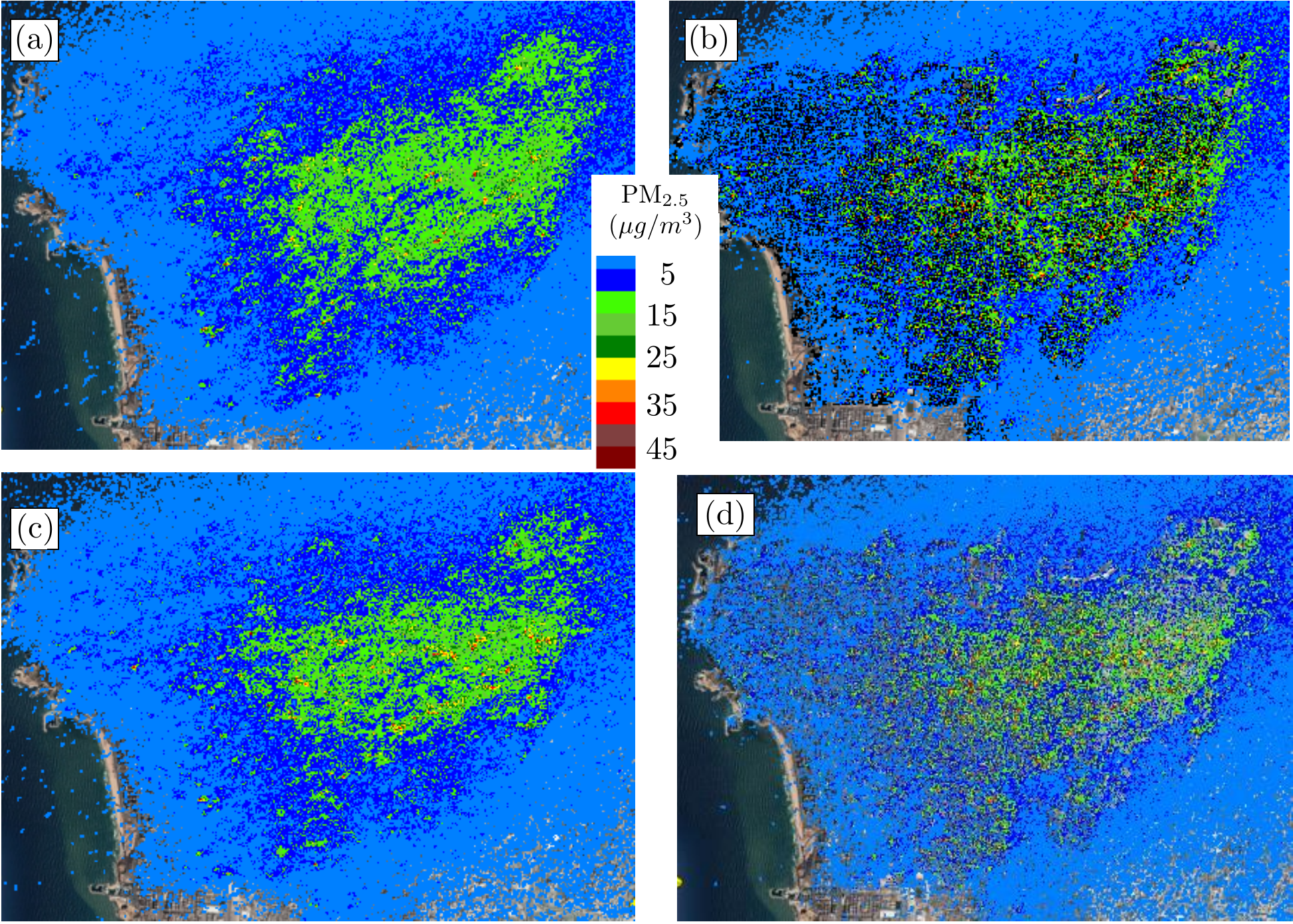}
		\caption{PM$_{2.5}$ concentration maps at 2 m above ground level for (a) no buildings, slightly unstable atmosphere (scenario 12),  (b) with buildings, slightly unstable atmosphere (scenario 13), (c) no buildings,  unstable atmosphere (scenario 10) and (d) with buildings, unstable atmosphere (scenario 11).}
		\label{FigPM2P5B1}
	\end{center}
\end{figure}

The impact of the atmospheric stability class can be discerned by comparing scenarios 13 and 11 (with buildings) and scenarios 12 and 10 (no buildings). In the presence of the buildings, the spatial distributions of the PM$_{2.5}$ concentration at 2 m above the ground for the cases of slightly unstable (scenario 13) and unstable (scenario 11) atmosphere are shown in Figs. \ref{FigPM2P5B1}-b and d at steady conditions. When comparing the two distributions, we notice that they are very similar, and the values of the normalized first floor and total exposure proxy indicators $\pi_1/\pi_1^*$ and $\Pi/\Pi^*$ are the same.\\
In the absence of buildings, the PM$_{2.5}$ concentration maps are shown in Figs. \ref{FigPM2P5B1}-a and c for the cases of slightly unstable (scenario 12) and unstable (scenario 10) atmosphere. When comparing the two distributions, we notice that for a slightly unstable atmosphere, the plume is slightly more stretched along the wind direction, exhibiting changes in the spatial variations in accordance with the observations made in section 3.1. The corresponding values of the normalized first floor and total exposure proxy indicators  ($\pi_1/\pi_1^*=0.78$ and $\Pi/\Pi^*=0.87$ for scenario 12 and $\pi_1/\pi_1^*=0.82$ and $\Pi/\Pi^*=0.83$ for scenario 10) are, however, close to each other. These results suggest that, while the atmospheric stability class changes the PM$_{2.5}$ concentration distribution locally, it weakly affects the total exposure for the given wind conditions and stack heights.  

\subsubsection{Effect of Wind Speed and Direction}
\begin{figure}[!htb]
	\begin{center}
		\includegraphics[width=\columnwidth]{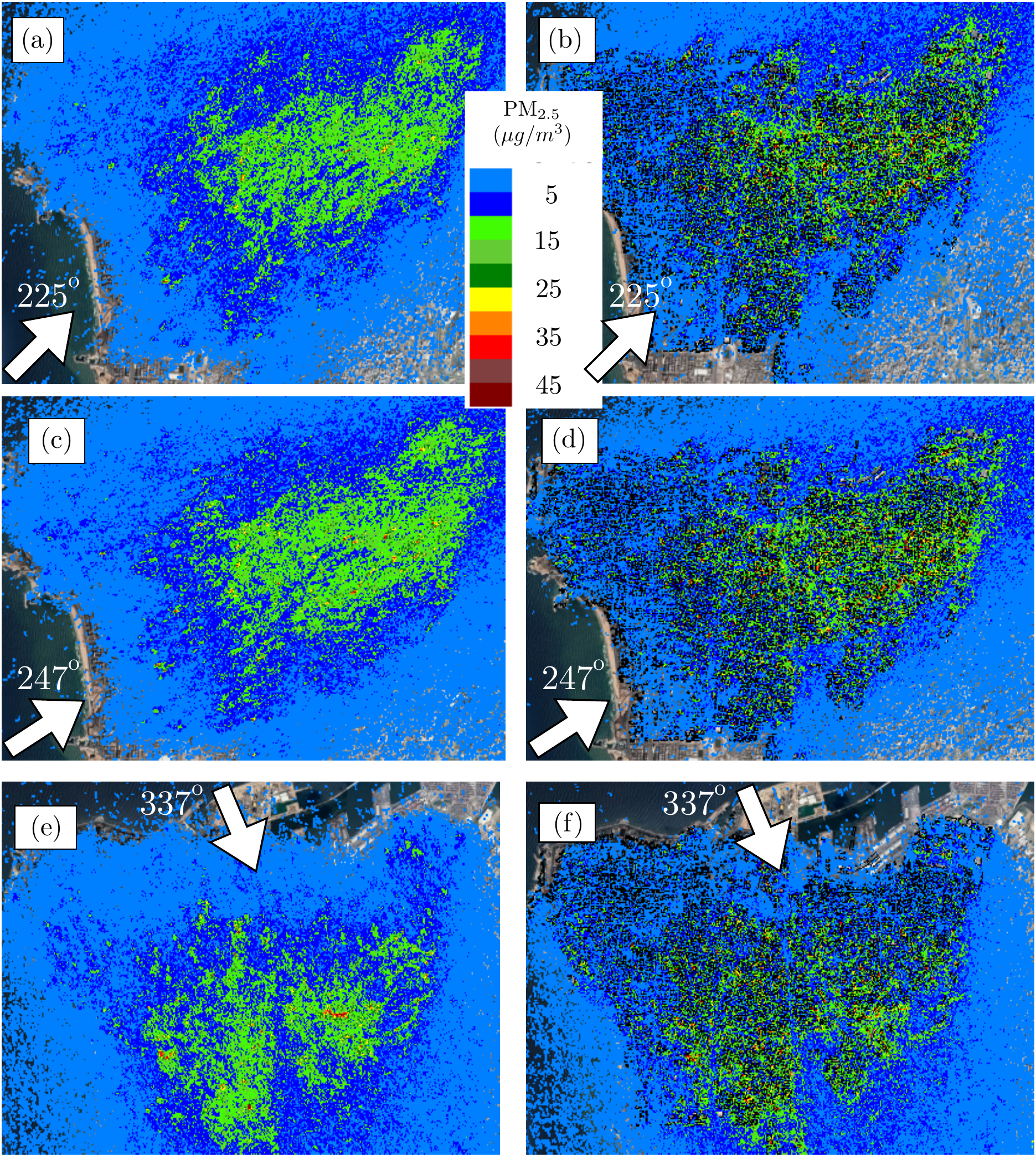}
		\caption{PM$_{2.5}$ concentration maps at 2 m above ground for (a) wind direction 225$^{\circ}$ without buildings (scenario 20) (b) wind direction 225$^{\circ}$ with buildings (scenario 21) (c) wind direction 247$^{\circ}$ without buildings (scenario 12) (d) wind direction 247$^{\circ}$ with buildings (scenario 13) (e) wind direction 337$^{\circ}$ without buildings (scenario 22) (f) wind direction 337$^{\circ}$ with buildings (scenario 23). Solutions are for a slightly unstable atmosphere at 4 m/s wind speed .}
		\label{Figwd}
	\end{center}
\end{figure}

The wind conditions are among the main meteorological conditions that affect the dispersion of PM$_{2.5}$ and other pollutants. The impact of the wind direction on the PM$_{2.5}$ map is shown in Figure \ref{Figwd} for the following prevailing wind directions: 225$^{\circ}$ (January - July and December),  247$^{\circ}$ (August and September), and 337$^{\circ}$ (October and November). The figures on the left are for scenarios without buildings and those on the right are for scenarios with buildings. As expected, we notice spatial variation of the pollutant dispersion along the direction of the downstream wind.  Since the wind direction differs by only 22$^{\circ}$, the large scale patterns in Figures  \ref{Figwd}-a and c (scenarios 20 and 12) and Figures  \ref{Figwd}-b and d (scenarios 21 and 13) are similar. For the scenarios (22 and 23) where the wind direction is 337$^{\circ}$, we notice a southward shift in the dispersion pattern with clear separation between and eastern and a western zone along the shallow valley between the two hills (see Figure \ref{FigBeirutTopStream}). The impact of the wind direction on the spatial distribution of the human exposure proxy indicator can be seen in Figures~\ref{Figwdexp}-a, b, and c for scenarios 21 (WD=225$^{\circ}$), 13  (WD=247$^{\circ}$) and 23 (WD=337$^{\circ}$). The distributions are over the latitude-longitude plane of Beirut at different elevations corresponding to the different floors of the buildings.By closely examining these maps, we notice that a change in the wind direction significantly changes the positions of the high human exposure zones. Yet, these human exposure hotspots are always shifted along the downstream wind in a similar manner to the shift of the pollutant dispersion pattern itself. \\
The values of the normalized first floor and total exposure proxy indicators for the considered wind directions, presented in Figs. \ref{barChartsAllAll}-a and e, show that the exposure is largest when the wind direction is 225$^\circ$, which may be explained by the fact that this wind direction is the closest to the direction orthogonal to the mountain range east of the city. We note that the values corresponding to the 337$^\circ$ wind direction  may not be as accurate since the population density distribution in the southern region of the city is not well represented. As for the normalized mean concentrations at 2 m elevation, plotted in Fig. \ref{barChartsAllAll}-c, they are roughly the same for all the scenarios without buildings (12, 20, 22)  with $\overline{PM}_{2.5}(2m) = 5.79 - 6.03 \, \mu g/m^3$ and for the scenarios  with buildings (13, 21, 23)  with $\overline{PM}_{2.5}(2m) = 7.22 - 7.45 \, \mu g/m^3$.
\begin{figure}[!htb]
	\begin{center}
		\includegraphics[width=\columnwidth]{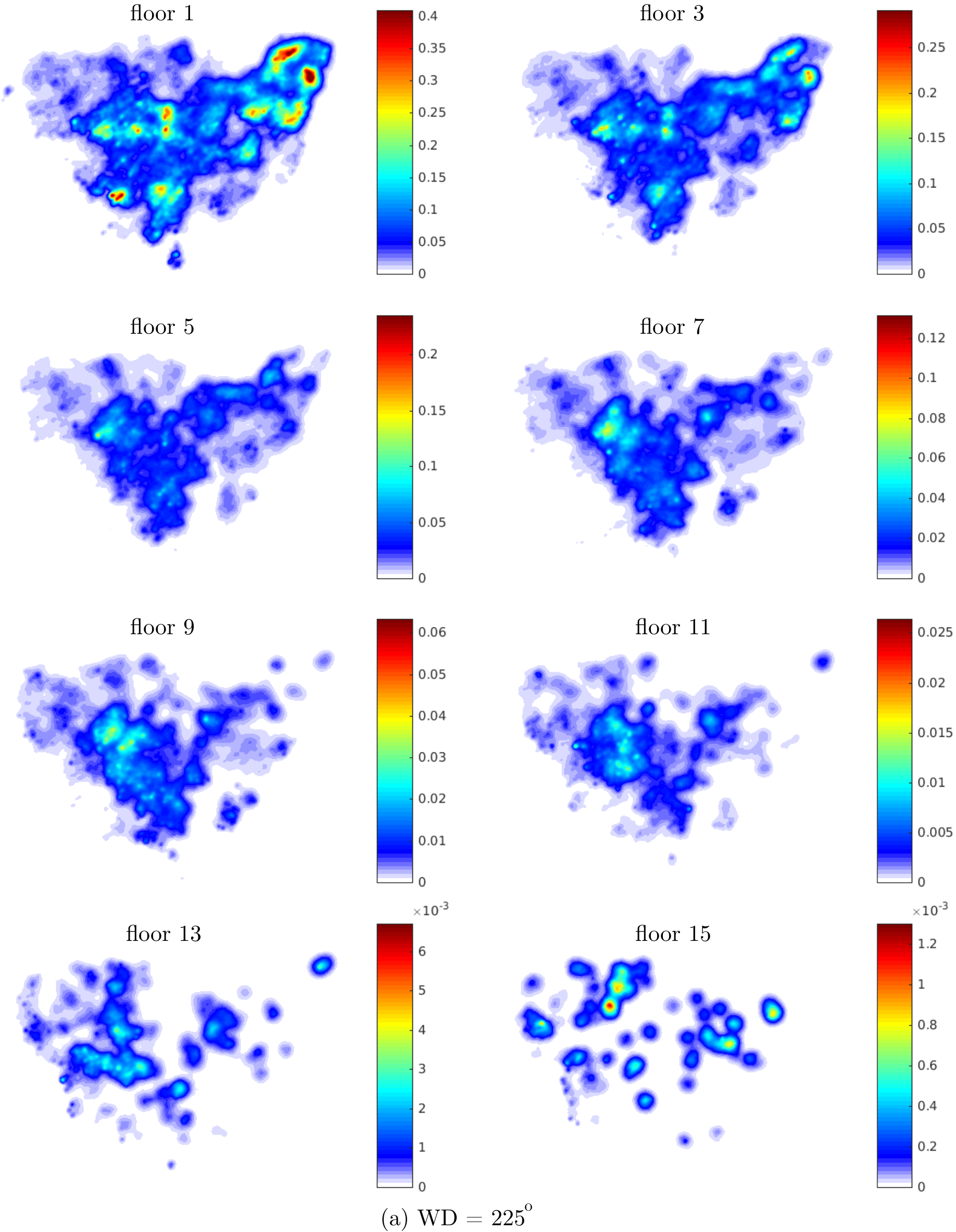}
	\end{center}
\end{figure}

\begin{figure}[!htb]
	\begin{center}
		\includegraphics[width=\columnwidth]{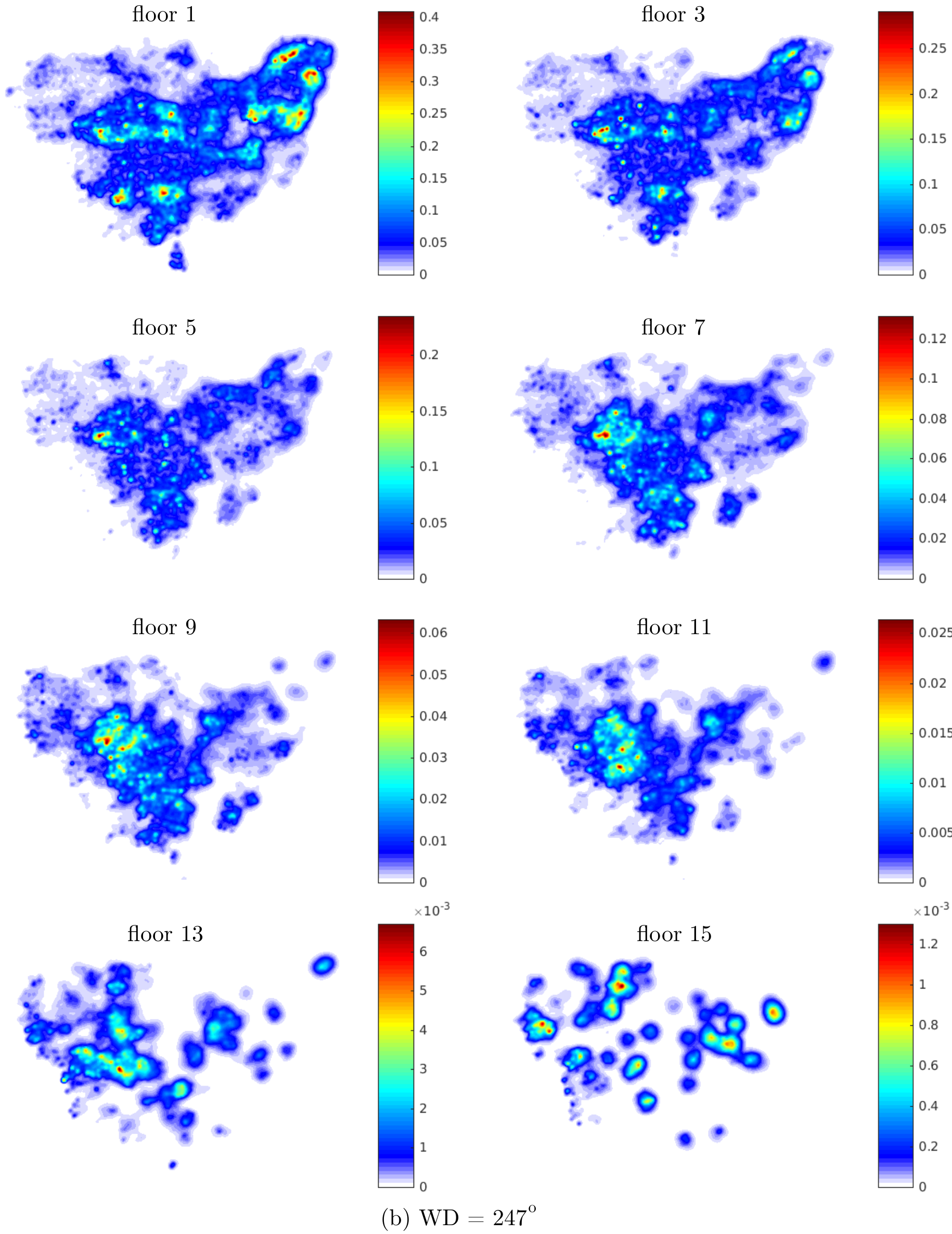}
	\end{center}
\end{figure}
\begin{figure}[!htb]
	\begin{center}
		\includegraphics[width=\columnwidth]{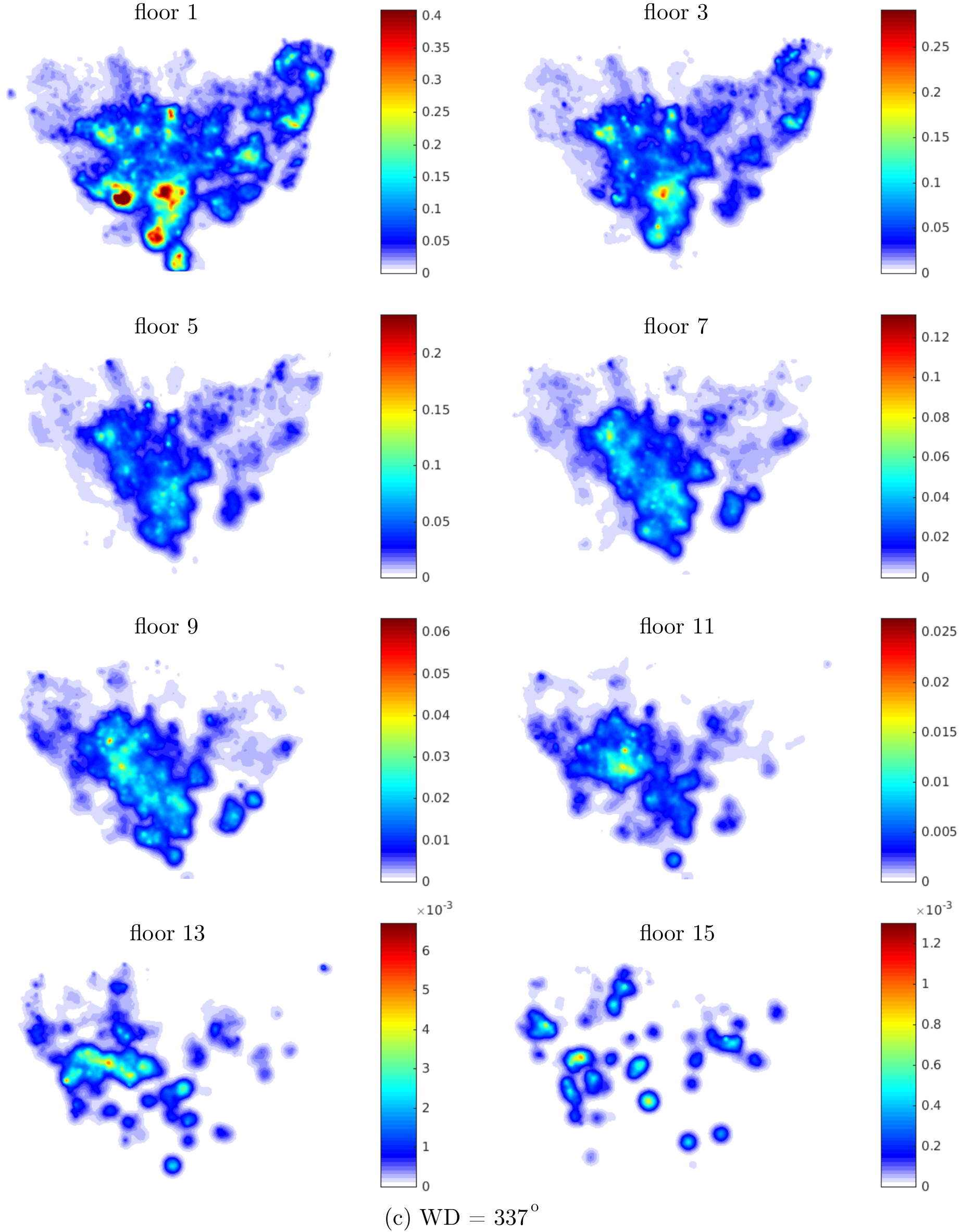}		
		\caption{Spatial distribution of the human exposure proxy indicators over Beirut for eight different floors for a wind speed of 4m/s, slightly unstable atmosphere, stack heights 2 meters above the roof tops. The prevailing wind directions are  a) 225$^{\circ}$ (scenario 21), b) 247$^{\circ}$ (scenario 13), and c) 337$^{\circ}$ (scenario 23).}
		\label{Figwdexp}
	\end{center}
\end{figure} 
\FloatBarrier

\begin{figure}[!htb]
	\begin{center}
		\includegraphics[width=4.5in]{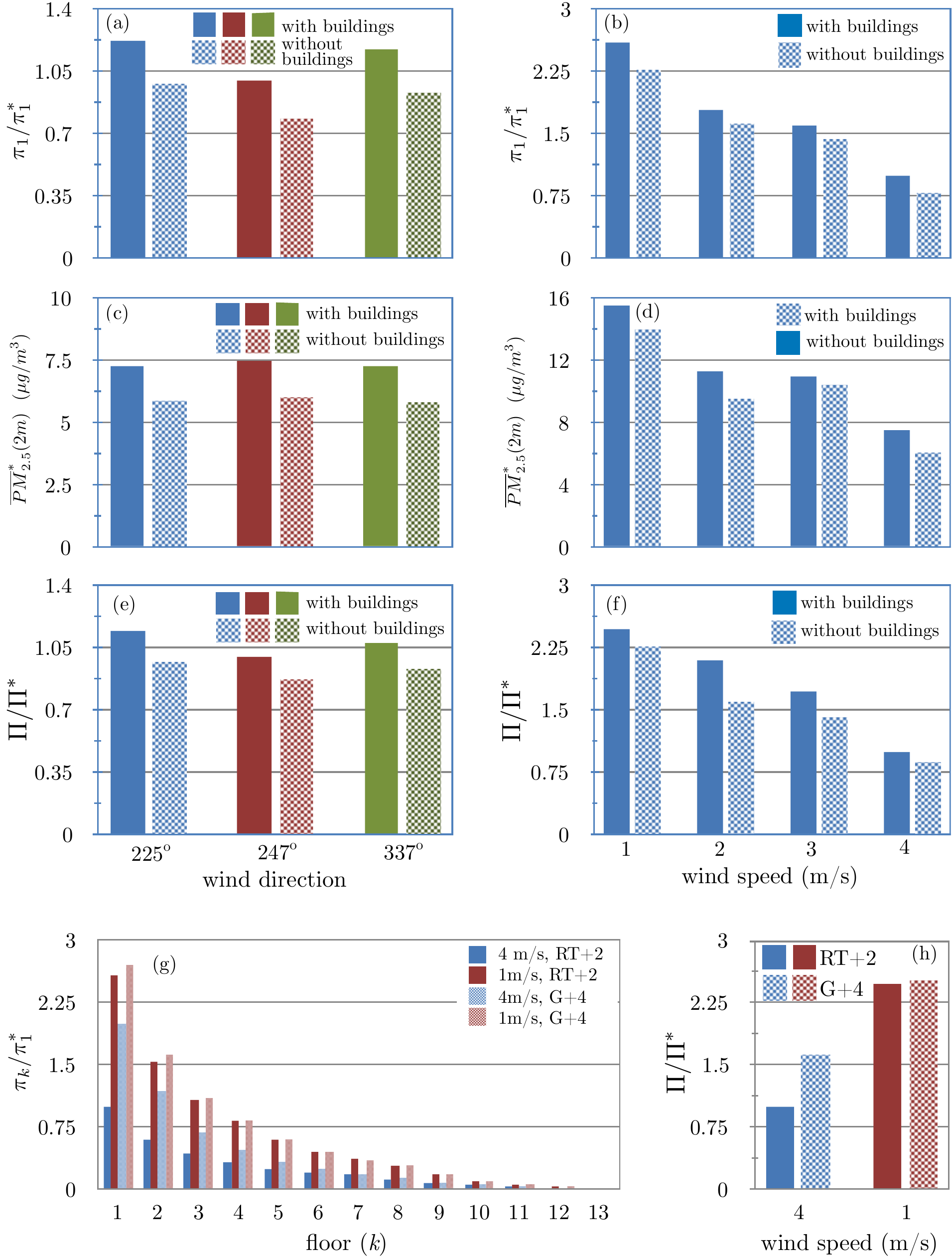}
	\end{center}
	\caption{(a) normalized first floor exposure proxy indicator versus wind direction, (b) normalized first floor exposure proxy indicator versus wind speed ,
		(c) mean ground level concentration  versus wind direction, (d) mean ground level concentration versus wind speed,
		(e) normalized total exposure proxy indicator versus wind direction, (f) normalized total exposure proxy indicator versus wind speed , 
		(g)  normalized floor exposure proxy indicator for  the different floors, and (h) Impact of wind speed and stack height on total exposure proxy indicator. Unless otherwise specified, the following conditions apply: slightly unstable atmosphere, wind direction: 247$^{\circ}$, wind speed: 4 m/s, stack heights: 2m above rooftops.}
	\label{barChartsAllAll}
\end{figure}

\begin{figure}[!htb]
	\begin{center}
		\includegraphics[width=4.3in]{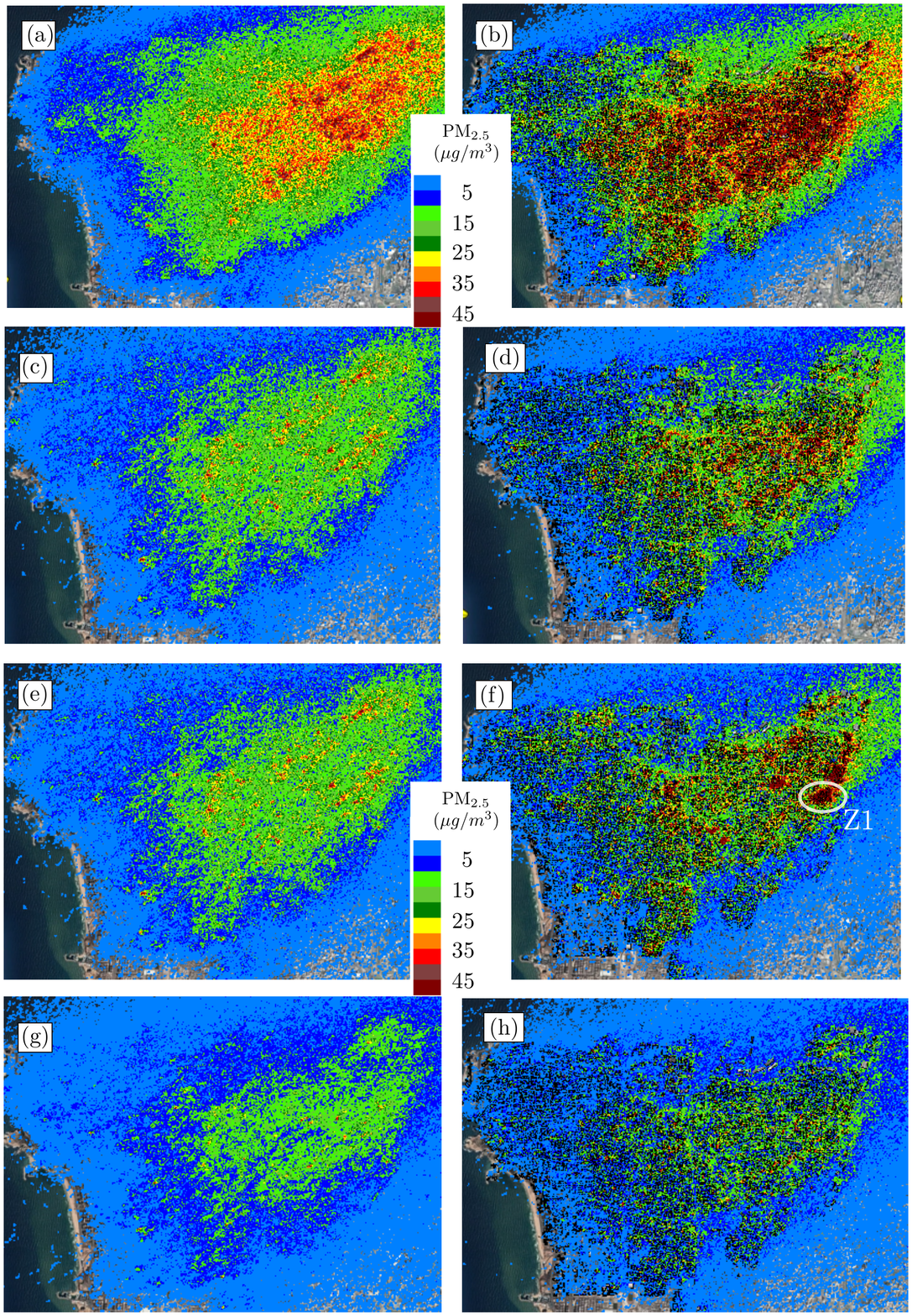}
		\caption{PM$_{2.5}$ concentration maps at 2 m above ground for (a) wind speed 1 m/s without buildings (scenario 28) (b) wind speed 1 m/s with buildings (scenario 29) (c) wind speed 2 m/s without buildings (scenario 26) (d) wind speed 2 m/s with buildings (scenario 27) (e) wind speed 3 m/s without buildings (scenario 24) (f) wind speed 3 m/s with buildings (scenario 25) (g) wind speed 4 m/s without buildings (scenario 12) (h) wind speed 4 m/s with buildings (scenario 13). Solutions are for a slightly unstable atmosphere and wind direction of 247$^{\circ}$.}
		\label{Figspd}
	\end{center}
\end{figure}

The impact of the wind speed on the PM$_{2.5}$ concentration map is shown in Figure \ref{Figspd} for the following wind speeds: 1 m/s (Figures \ref{Figspd}-a and b), 2 m/s (Figures \ref{Figspd}-c and d), 3 m/s (Figures \ref{Figspd}-e and f), and 4 m/s (Figures \ref{Figspd}-g and h). These predictions are for a slightly unstable atmosphere and the prevailing wind direction of 247$^{\circ}$.
The figures on the left are for the scenarios without buildings and those on the right are for the scenarios with buildings. We notice that as the wind speed decreases, both the intensity and the spatial coverage of the PM$_{2.5}$ concentration distribution increases. For the considered prevailing wind direction, the impact of the wind speed on the spatial distribution of the human exposure proxy indicator can be seen in Figures~\ref{Figspdexp}-a, b, c and d for scenarios 29 (WS=1 m/s), 27  (WS = 2 m/s), 25 (WS = 3 m/s) and 13 (WS = 4 m/s). The distributions are over the latitude-longitude plane of Beirut at different elevations corresponding to the different floors of the buildings. By examining the figures, we notice that the distribution is largely insensitive to the prevailing wind speed over the considered range (1-4 m/s), where it is observed that the locations of the human exposure hot spots do not vary for the considered wind direction. The intensity of the human exposure proxy indicator is, however, strongly dependent on the wind speed, with larger intensities observed for lower wind speeds. The impact of the wind speed on the first floor and total exposure proxy indicators and on $\overline{PM}_{2.5}(2m)$ is shown in Figs. \ref{barChartsAllAll}-b, f and d respectively. As the speed decreases from 4 m/s to 1 m/s, the respective increases in 
$\pi_1$, $\Pi$, and $\overline{PM}_{2.5}(2m)$ are 259 \%, 247\% and 208\% in the presence of buildings. The difference in percent increases between $\pi_1$ and $\overline{PM}_{2.5}$ is attributed to the fact that the population density is higher in regions of higher PM$_{2.5}$ concentration. \\
For low wind speeds ($\leq $3 m/s), localized highly polluted zones (similar to zone Z1 observed in scenarios 25  of Fig. \ref{Figspd}-f) 
start to appear, especially for the scenarios where the buildings are present (Figs.  \ref{Figspd}-b, d, f, and h). The number and intensity of these zones increase as the wind speed decreases. These observations are a reflection of the impact of the wind speed, which cannot be overstated. Higher wind speeds enhance the ventilation of pollutants while lower wind speeds promote accumulation of pollutants. Low wind speeds, together with short street canyons that are oriented perpendicular to the incoming wind direction are the ingredients that make up highly polluted zones. Notice that for the case of diesel generators, short street canyons imply that the emission sources (within the canyon) are necessarily located near the ground.

\begin{figure}[!htb]
	\begin{center}
		\includegraphics[width=\columnwidth]{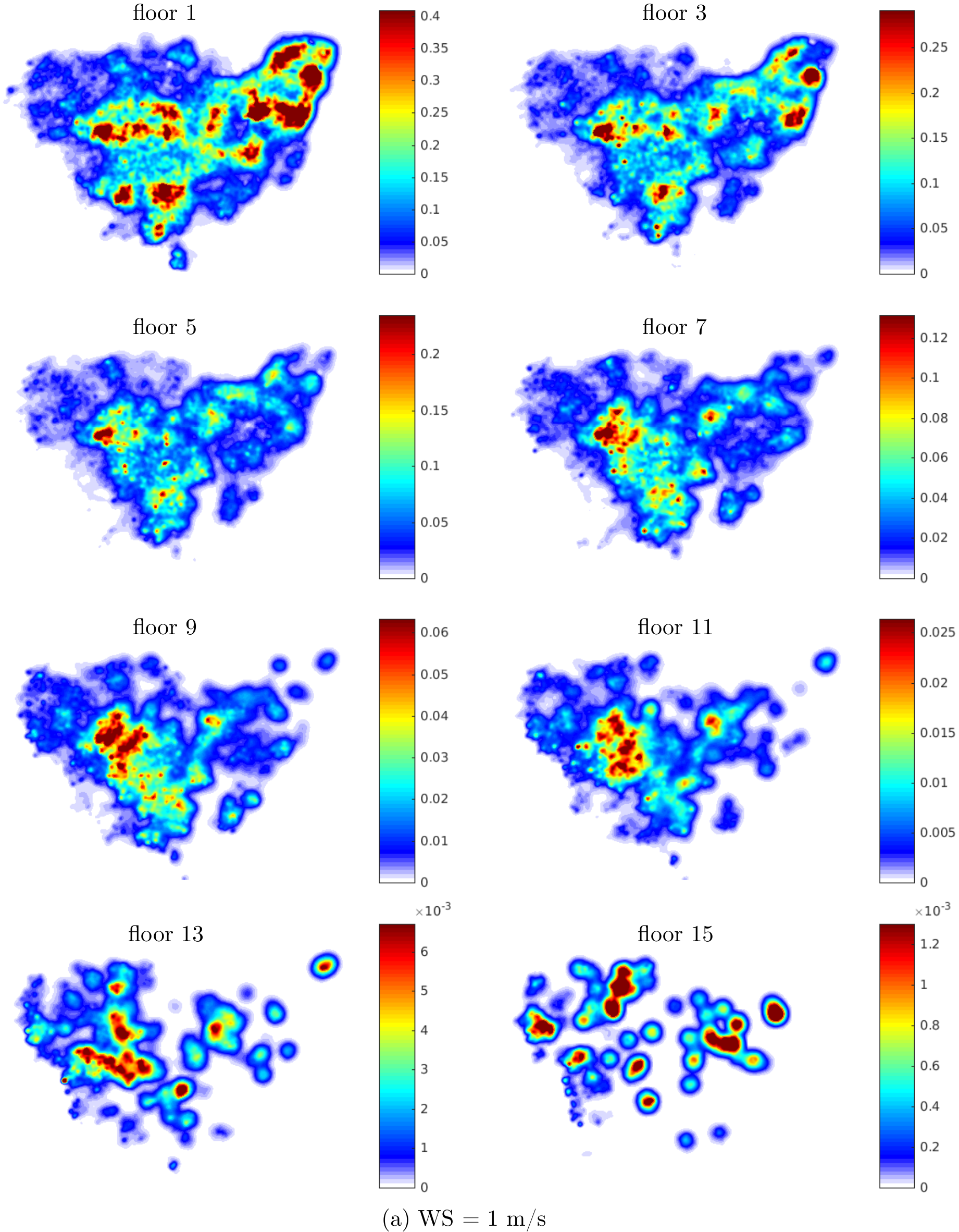}
	\end{center}
\end{figure}
\begin{figure}[!htb]
	\begin{center}
		\includegraphics[width=\columnwidth]{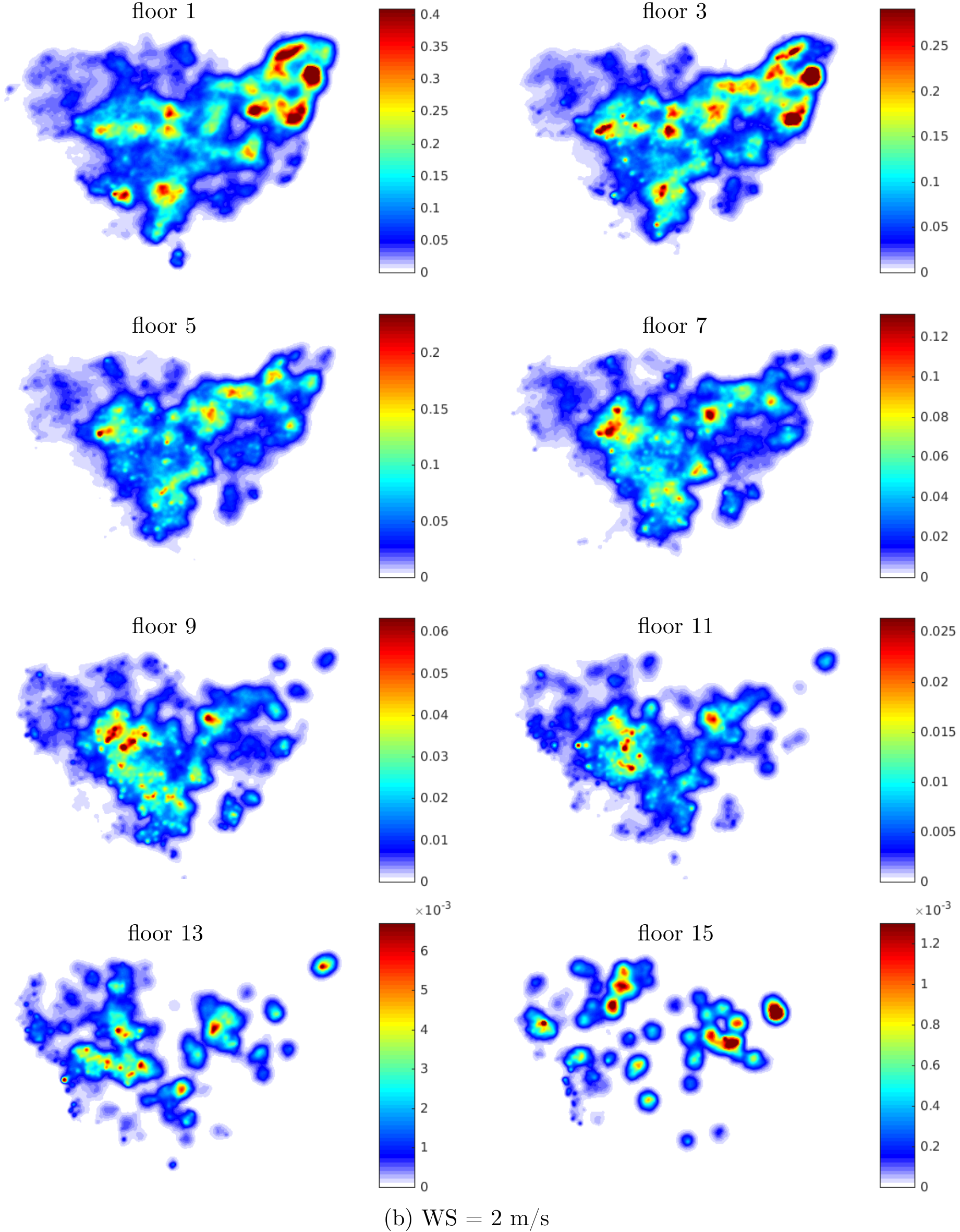}
	\end{center}
\end{figure}
\begin{figure}[!htb]
	\begin{center}
		\includegraphics[width=\columnwidth]{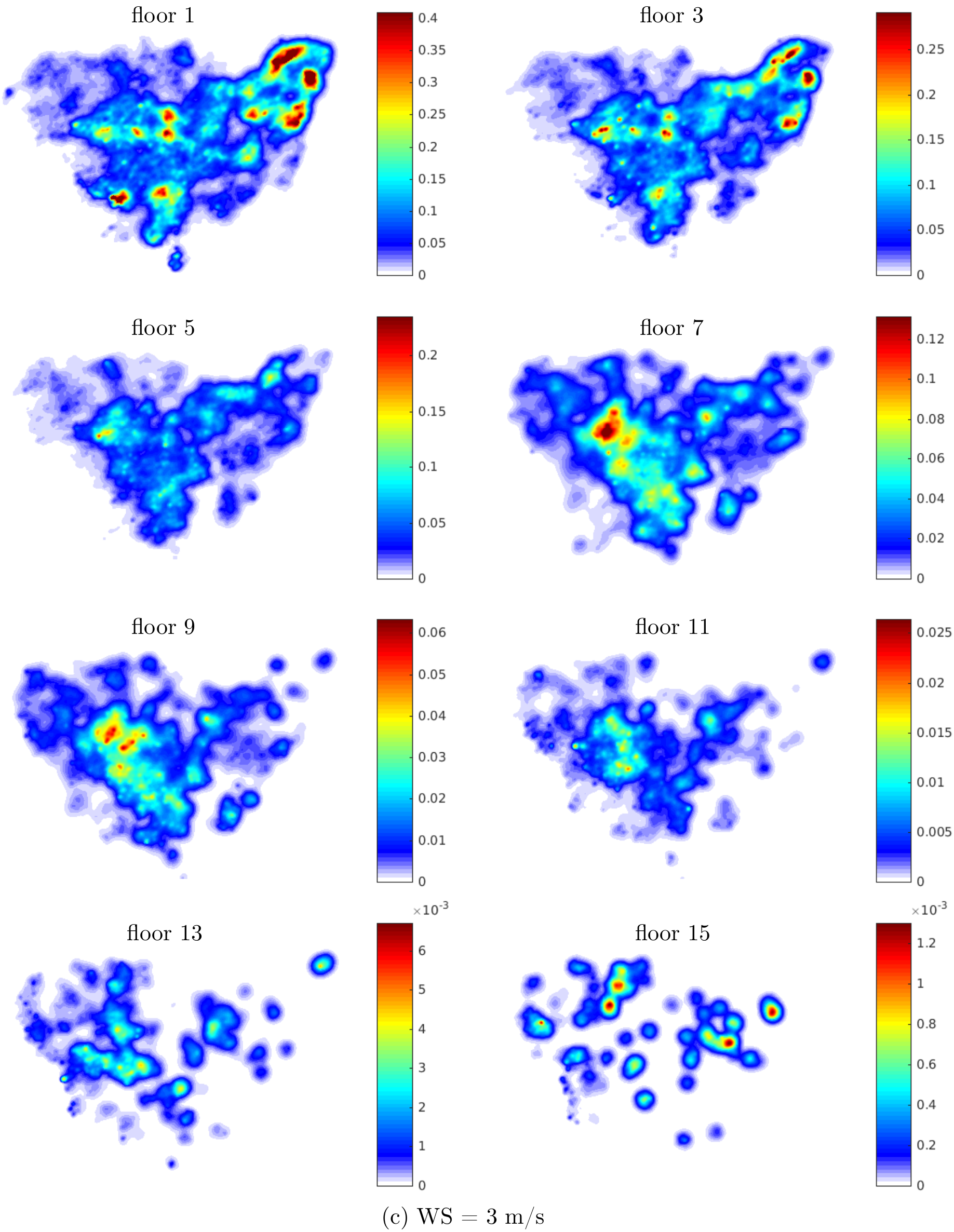}
	\end{center}
\end{figure}
\begin{figure}[!htb]
	\begin{center}
		\includegraphics[width=\columnwidth]{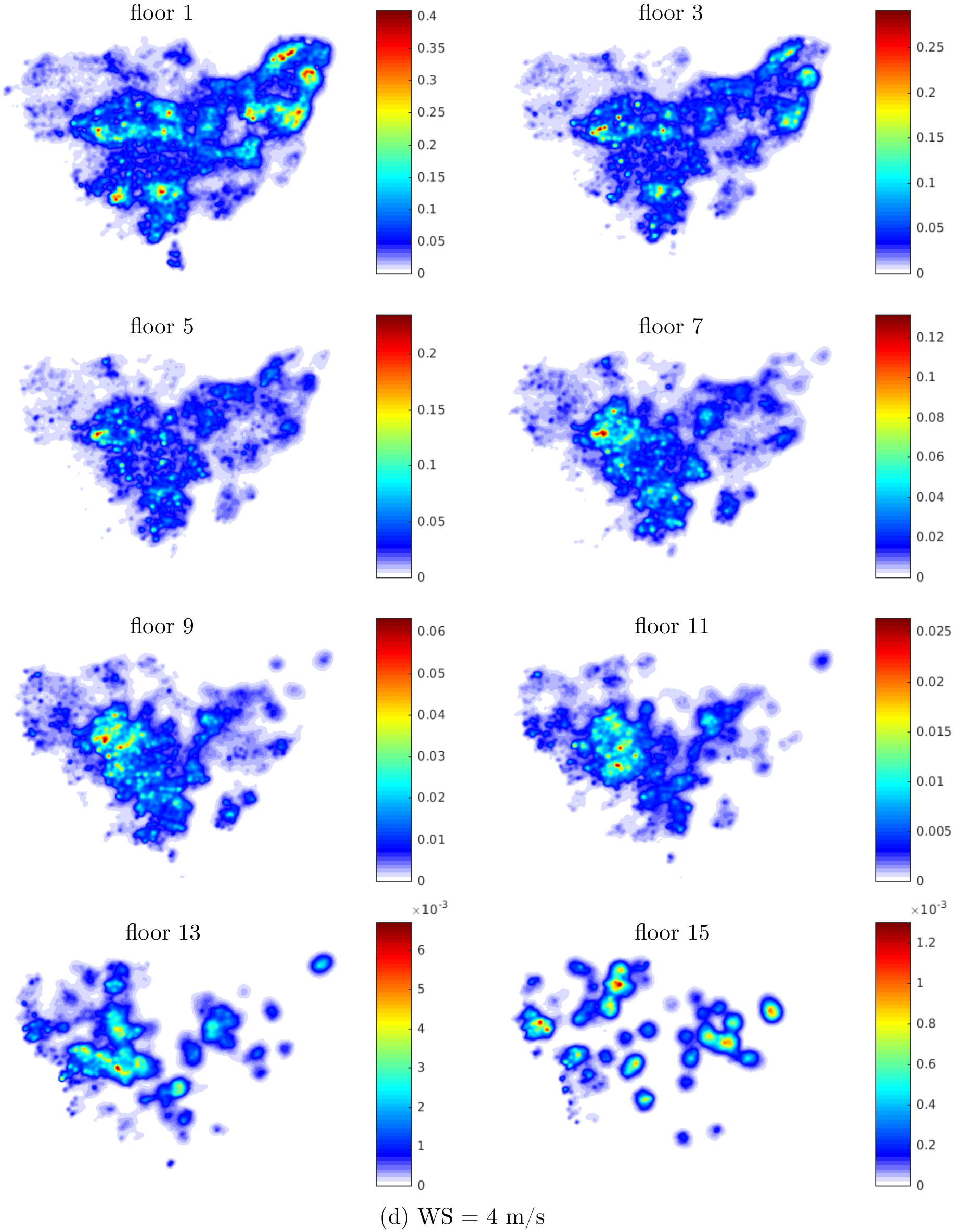}		
		\caption{Spatial distribution of the human exposure proxy indicators over Beirut for eight different floors for a wind direction of 247$^{\circ}$, slightly unstable atmosphere, stack heights 2 meters above the roof tops. The prevailing wind directions are a) 1m/s (scenario 29), b) 2m/s (scenario 27), c) 3m/s (scenario 25) and d) 4m/s (scenario 13).}
		\label{Figspdexp}
	\end{center}
\end{figure} 
\FloatBarrier

\subsubsection{Effect of Stack Heights}
\begin{figure}[!htb]
	\begin{center}
		\includegraphics[width=\columnwidth]{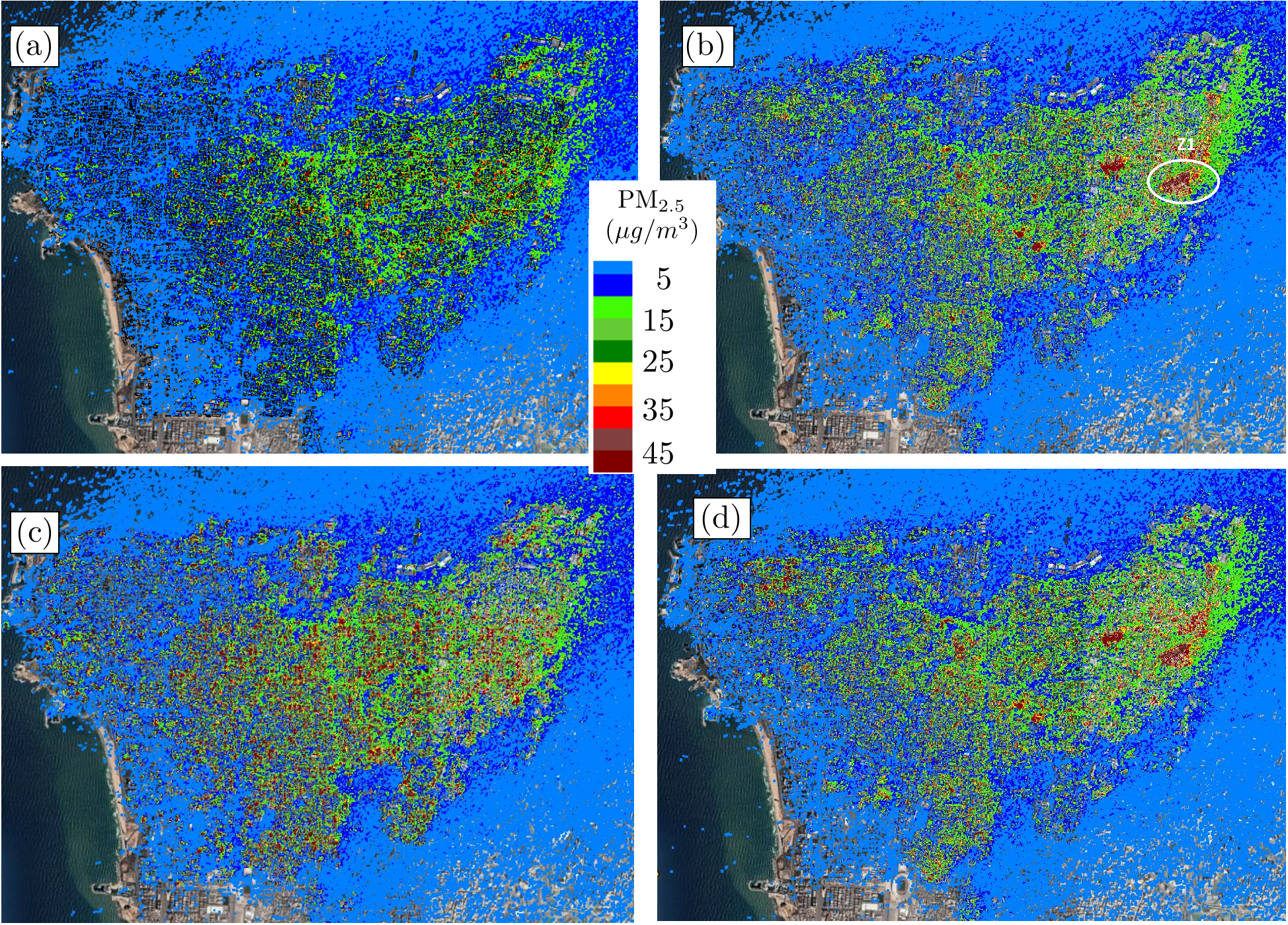}
		\caption{PM$_{2.5}$ concentration maps at 2 m above ground for (a) scenario 13,  (b) scenario 17, (c)  scenario 18, and (d) scenario 19. Solutions are for a slightly unstable atmosphere.}
		\label{FigPM2P5B2}
	\end{center}
\end{figure}

Figures \ref{FigPM2P5B2}-a and c show the spatial distributions PM$_{2.5}$ concentration at 2 m above the ground for scenarios 13 and 18. While these two scenarios share the same wind speed (4 m/s) and direction (225$^\circ$), atmospheric stability class (slightly unstable), generators distribution in the longitude-latitude plane (GD1), they differ in the stack heights. Upon placing the emission sources at the street level (4 m above ground), the map shown in Fig. \ref{FigPM2P5B2}-c shows that a larger area is affected with the presence of many more hot spots as compared to the case when the stacks are at 2 m above the rooftops (Fig. \ref{FigPM2P5B2}-a).
Accordingly, the first floor proxy indicator, the total exposure proxy indicators, and the  mean PM$_{2.5}$ concentration at 2 m increase by 99\%, 63\% and 70\% when the emissions source are brought down to the street level.\\
Similar observations can be made for corresponding two scenarios using generators distribution GD2,  which randomly allocates one generator per two buildings (refer to section \ref{SubsecDomain} for more details). The PM$_{2.5}$ concentration  maps of scenarios 17 and 19 are shown in Figures. \ref{FigPM2P5B2}-b and d.

\begin{figure}[!htb]
	\begin{center}
		\includegraphics[width=\columnwidth]{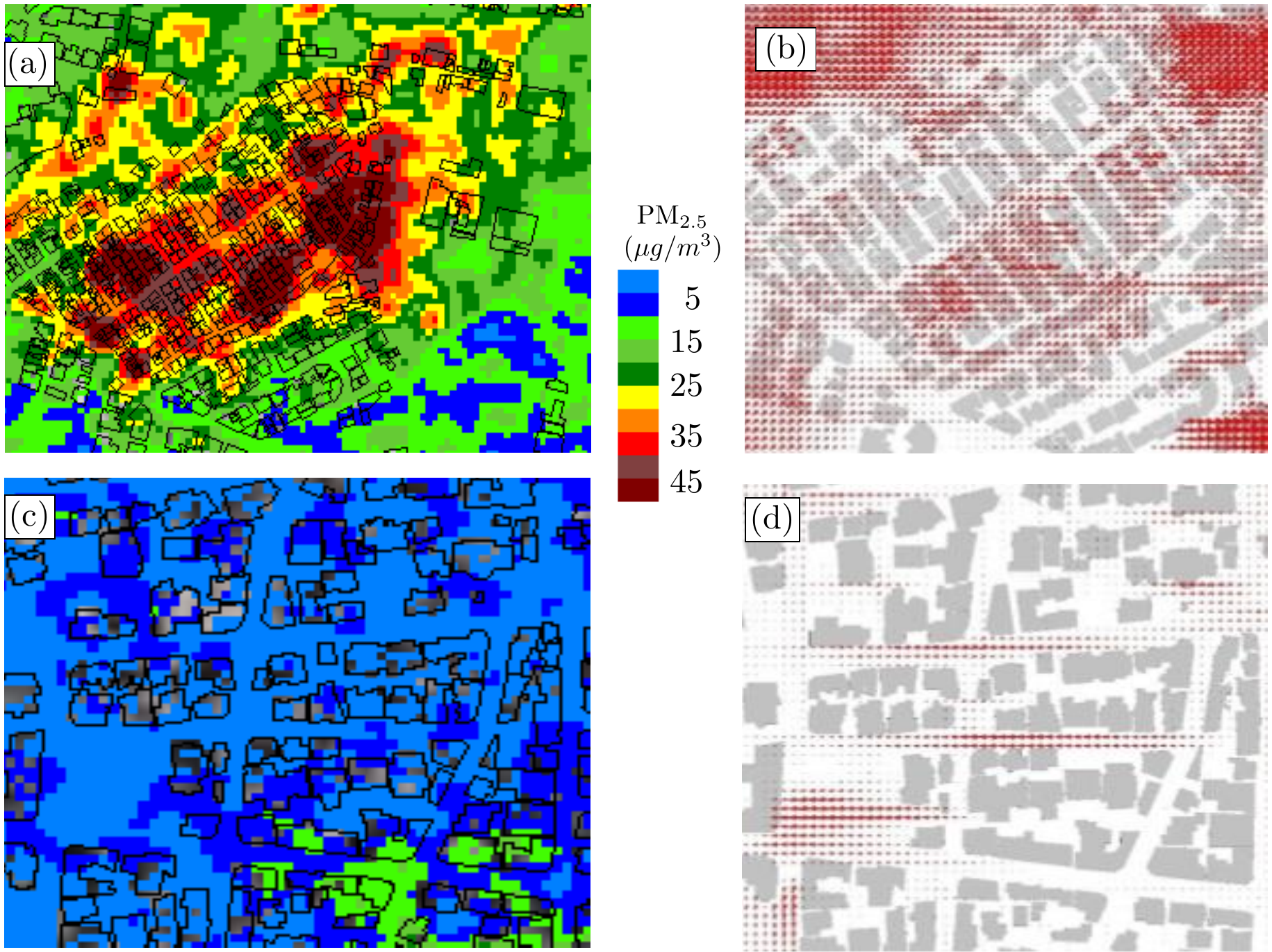}
		\caption{(left) PM$_{2.5}$ concentrations at 2 m above ground leveland (right) wind at roof level (19 m) in two different street canyons with orientations: (a,b) perpendicular to the wind flow and (c,d) parallel to the wind flow .}
		\label{Fig14}
	\end{center}
\end{figure}

Lowering down the stacks from roof tops to street level does not, however, has as much impact at lower wind speeds. As can be observed in Fig.  \ref{barChartsAllAll}-h, lowering the stack to 4 m above the ground results in less than 5\% increase in both the first floor and the total exposure proxy indicators when the wind speed is 1 m/s, as opposed to an increase exceeding 60\% when the wind speed is 4 m/s. Note for all of these cases, we consider a slightly unstable atmosphere with a prevailing wind direction of 247$^{\circ}$ in the presence of buildings. Fig.  \ref{barChartsAllAll}-g shows a plot of the normalized fist floor proxy indicator versus floor number for these four scenarios (13, 18, wind speeds of 4 and 1 m/s and stack heights of  2 meters above the roof top (scenarios 13 and 29) and 
4 meters above the ground level (scenarios 18 and 30). The plot shows that, for wind speed of 4 m/s, moving the stacks from street level to above roof tops significantly reduces the exposure of residents of all floors. This reduction reaches nearly 50\% for the first two floors. This is not the case, however, for wind speed of 1 m/s, where it can be seen that moving the stacks to the roof tops yield does not reduce the exposure. One can also see that the general trend is that of a smaller population weighted integrated concentration for higher floors, which is expected given the floors distribution among the buildings; i.e. lower floors have more inhabitants. By summing up the total exposure of all floors, we notice it is always higher for the scenarios when the stacks are placed on the ground level. Furthermore, when the prevailing wind speed decreases, the exposure becomes less sensitive to the stacks height as shown in Fig.  \ref{barChartsAllAll}-h. Thus increasing the stack height is an effective mitigation measure only at high enough wind speeds. \\
Upon inspecting Figs. \ref{FigPM2P5B2}-a and b, one can observe that the two generators distributions (GD1 and GD2) yield similar large-scale patterns of PM$_{2.5}$ dispersion, implying that the dispersion pattern is, to a large degree, insensitive to the exact location of the stacks, subject to the constraints governing the underlying statistical distributions of their longitude-latitude locations. Upon placing the emission sources at the street level (4 m above ground), the maps shown in Figures \ref{FigPM2P5B2}-c and d, exhibit similar patterns on the large scale. One can, however, observe the existence of localized highly polluted zones (e.g. zone Z1) in scenarios 17 and 19 (figure \ref{FigPM2P5B2}-b and d). Comparison with figures \ref{FigPM2P5B2}-a and c shows that  the pollution dispersion in such zones is highly sensitive to the locations of the emission sources and their emission rates. These zones are commonly found in regions of low velocity, high buildings density, and street canyons perpendicular to the wind direction. In Zone Z1, the canyon is oriented perpendicular to the wind direction, pockets of pollution are observed (Figures \ref{Fig14}-a and b).  According to the ratios of building height to street width and building length to height, these canyons are classified as deep narrow canyons where the most circular eddy emerges in the skimming flow regime \cite{oke1988street}. These observations are in line with what is reported in the literature under similar conditions \cite{afiq2012effects,yazid2014review}. 
On the other hand, when  the street canyon is parallel to the predominant wind direction, the wind is accelerated through the canyon driving the pollution away from the canyon (Figures \ref{Fig14}-c and d).

\begin{figure}[!htb]
	\begin{center}
		\includegraphics{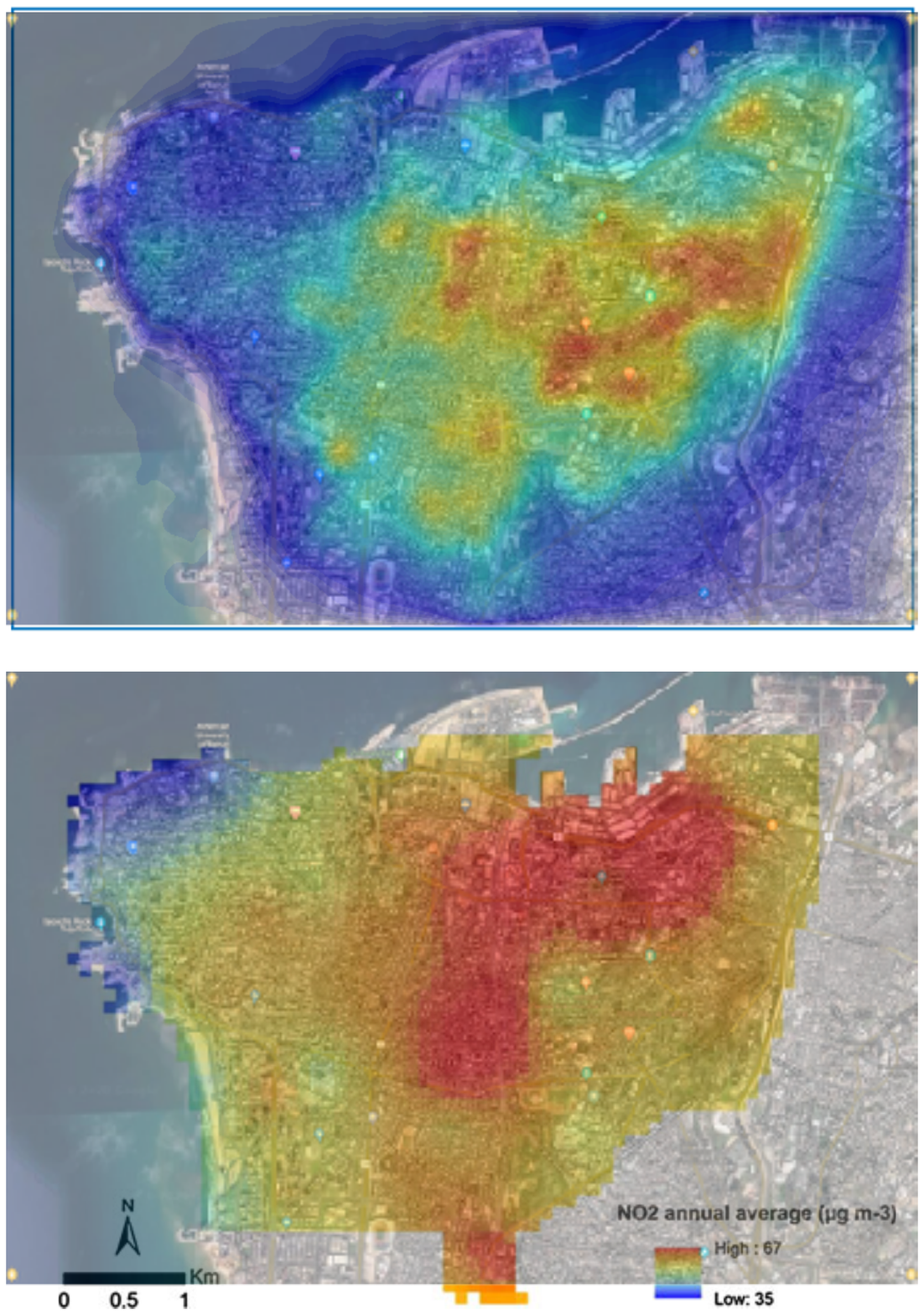}
		\caption{Comparison of the the model predicted PM$_{2.5}$ concentration map (top) with the $NO_2$ map (bottom) reported in ~\cite{badaro2014geostatistical}.}
		\label{comparisonWithBadaroSaliba}
	\end{center}
\end{figure}

We conclude by comparing the model predicted PM$_{2.5}$ concentration map at street level with the NO$_2$ map reported in~\cite{badaro2014geostatistical}. The NO$_2$ map reported in~\cite{badaro2014geostatistical} is based on measurements carried over seven months. The average monthly wind directions over these months are: 247$^\circ$ (case 13) for the months of February, March, June and July,  337$^\circ$ (case 23) for the months of months od October and November, and 225$^\circ$ (case 21) for the month of December. The corresponding weighted average of the PM$_{2.5}$ maps (at 2m above ground) of cases 13, 23 and 21 is shown at the top of figure~\ref{comparisonWithBadaroSaliba}. Similarity in the pattern and spatial coverage can be observed when compared with the  NO$_2$ map reported in~\cite{badaro2014geostatistical}, shown at the bottom of the figure.

\section{Conclusion and Future Work}\label{SecConc}
In this work, we carry out a model-based study of PM$_{2.5}$ dispersion originating  from a large number of diesel generators installed in Beirut. In particular, the spatial distribution of the PM$_{2.5}$ concentration, predicted at steady state using the computational modeling tool GRAMM-GRAL for Lagrangian transport, is utilized to explore large and small scale dispersion patterns in selected smalls domains and over the entire city. A number of scenarios were considered to investigate the impact of topography, atmospheric stability, presence of buildings, diesel generators distribution, and emission sources elevations (stacks heights) for representative meteorological conditions of the wind speed and direction.\\
The simulations in the small domains revealed a strong correlation between emission sources locations and emission rates, atmospheric stability, buildings geometries and orientations, and wind direction and speed. This renders simple mitigating measures, such as increasing stack height or changing its location, of limited effectiveness. We also found out the pollution hot spots that emerge in regions of low velocity, high density of short buildings , and street canyons perpendicular to the wind direction. The local dispersion patterns within these highly polluted zones are highly sensitive to the locations of the emission sources and their emission rates. We also found out that persistent cold spots are characterized by street canyons that are parallel to the predominant wind direction, the wind is accelerated through the canyon driving the pollution away from the canyon.\\
The simulations over the large domain showed that the large scale dispersion patterns are largely influenced by the wind speed and direction. While the wind direction shift the entire distribution along the wind direction, smaller wind speeds increases the intensity and expands the spatial coverage of the PM$_{2.5}$ concentration. For the case where the stacks are located at 2m above rooftops and for wind direction of 247$^\circ$, the simulations showed that reducing the wind speed from 4 m/s to 1 m/s resulted in more than two fold increase in the mean concentration at street level and the first floor and total exposure proxy indicators.  For a given wind direction, speed, and atmospheric stability condition, moving the stacks from street level to the top of the buildings results in significant reduction in the ground level concentration distribution at high speeds. At lower speeds, this reduction is less which limits the effectiveness of elevating the stacks to rooftops in mitigating street level exposure.\\
Impact of wind speed and stack height on the exposure proxy indicators for residents of floors 1 through 13 shows that for all cases the value of the indicator monotonically decreases for higher floors, as dictated by the vertical distribution of the population. For wind speed of 4 m/s, moving the stacks from street level to 2m above roof tops significantly reduces the exposure of residents of all floors. This reduction reaches nearly 50\% for the first two floors. This is not the case, however, for wind speed of 1 m/s, where it can be seen that moving the stacks to the roof tops yield does not reduce the exposure. One can also see that the general trend is that of a smaller population weighted integrated concentration for higher floors, which is expected given the floors distribution among the buildings; i.e. lower floors have more inhabitants.\\
By taking into account wind conditions, topography, buildings geometries, emission sources locations and emission rates, and population density distribution, We expect this model-based physical approach to be instrumental in assessing the impact on population exposure of what-if scenarios encountered in a spectrum of inquires ranging from epidemiological studies to urban planning.

\section{Acknowledgments}\label{Acknowledgments}
The authors acknowledge Dietmar Oettl for his support on using GRAMM-GRAL. They also acknowledge the support from the Collaborative Research Stimulus (CRS) at the American University of Beirut.

\section{Funding}\label{funding}
This work was funded by the American University of Beirut,  CRS Award $\#$ 103321.

\section*{}

\bibliography{biblio}

\begin{thebibliography}{10}
\expandafter\ifx\csname url\endcsname\relax
  \def\url#1{\texttt{#1}}\fi
\expandafter\ifx\csname urlprefix\endcsname\relax\def\urlprefix{URL }\fi
\expandafter\ifx\csname href\endcsname\relax
  \def\href#1#2{#2} \def\path#1{#1}\fi

\bibitem{du2019does}
Y.~Du, Q.~Wan, H.~Liu, H.~Liu, K.~Kapsar, J.~Peng, How does urbanization
  influence pm2. 5 concentrations? perspective of spillover effect of
  multi-dimensional urbanization impact, Journal of Cleaner Production 220
  (2019) 974--983.

\bibitem{xian2019recent}
C.~Xian, X.~Zhang, J.~Zhang, Y.~Fan, H.~Zheng, J.~Salzman, Z.~Ouyang, Recent
  patterns of anthropogenic reactive nitrogen emissions with urbanization in
  china: Dynamics, major problems, and potential solutions, Science of the
  Total Environment 656 (2019) 1071--1081.

\bibitem{world2016ambient}
W.~H. Organizations,
  \href{https:www.who.int/phe/health\_topics/outdoorair/databases/cities/en/}{Global
  urban ambient air pollution database}.
\newline\urlprefix\url{https:www.who.int/phe/health\_topics/outdoorair/databases/cities/en/}

\bibitem{brunekreef2002air}
B.~Brunekreef, S.~T. Holgate, Air pollution and health, The lancet 360~(9341)
  (2002) 1233--1242.

\bibitem{kappos2004health}
A.~D. Kappos, P.~Bruckmann, T.~Eikmann, N.~Englert, U.~Heinrich, P.~H{\"o}ppe,
  E.~Koch, G.~H. Krause, W.~G. Kreyling, K.~Rauchfuss, et~al., Health effects
  of particles in ambient air, International journal of hygiene and
  environmental health 207~(4) (2004) 399--407.

\bibitem{schwartz1996daily}
J.~Schwartz, D.~W. Dockery, L.~M. Neas, Is daily mortality associated
  specifically with fine particles?, Journal of the Air \& Waste Management
  Association 46~(10) (1996) 927--939.

\bibitem{atkinson2014epidemiological}
R.~Atkinson, S.~Kang, H.~Anderson, I.~Mills, H.~Walton, Epidemiological time
  series studies of pm2. 5 and daily mortality and hospital admissions: a
  systematic review and meta-analysis, Thorax 69~(7) (2014) 660--665.

\bibitem{dockery1993association}
D.~W. Dockery, C.~A. Pope, X.~Xu, J.~D. Spengler, J.~H. Ware, M.~E. Fay, B.~G.
  Ferris~Jr, F.~E. Speizer, An association between air pollution and mortality
  in six us cities, New England journal of medicine 329~(24) (1993) 1753--1759.

\bibitem{fann2012estimating}
N.~Fann, A.~D. Lamson, S.~C. Anenberg, K.~Wesson, D.~Risley, B.~J. Hubbell,
  Estimating the national public health burden associated with exposure to
  ambient pm2. 5 and ozone, Risk Analysis: An International Journal 32~(1)
  (2012) 81--95.

\bibitem{wilson1997fine}
W.~E. Wilson, H.~H. Suh, Fine particles and coarse particles: concentration
  relationships relevant to epidemiologic studies, Journal of the Air \& Waste
  Management Association 47~(12) (1997) 1238--1249.

\bibitem{abdallah2018first}
C.~Abdallah, C.~Afif, N.~El~Masri, F.~{\"O}zt{\"u}rk, M.~Kele{\c{s}},
  K.~Sartelet, A first annual assessment of air quality modeling over lebanon
  using wrf/polyphemus, Atmospheric Pollution Research 9~(4) (2018) 643--654.

\bibitem{liu2004large}
C.-H. Liu, M.~C. Barth, D.~Y. Leung, Large-eddy simulation of flow and
  pollutant transport in street canyons of different
  building-height-to-street-width ratios, Journal of Applied Meteorology
  43~(10) (2004) 1410--1424.

\bibitem{heist2013estimating}
D.~Heist, V.~Isakov, S.~Perry, M.~Snyder, A.~Venkatram, C.~Hood, J.~Stocker,
  D.~Carruthers, S.~Arunachalam, R.~C. Owen, Estimating near-road pollutant
  dispersion: A model inter-comparison, Transportation Research Part D:
  Transport and Environment 25 (2013) 93--105.

\bibitem{abdel2008atmospheric}
A.~A. Abdel-Rahman, On the atmospheric dispersion and gaussian plume model, in:
  Wwai'08: Proceedings of the 2nd International Conference on Waste Management,
  Water Pollution, Air Pollution, Indoor Climate, 2008, pp. 31--39.

\bibitem{di2008flow}
S.~Di~Sabatino, R.~Buccolieri, B.~Pulvirenti, R.~Britter, Flow and pollutant
  dispersion in street canyons using fluent and adms-urban, Environmental
  Modeling \& Assessment 13~(3) (2008) 369--381.

\bibitem{pullen2005comparison}
J.~Pullen, J.~P. Boris, T.~Young, G.~Patnaik, J.~Iselin, A comparison of
  contaminant plume statistics from a gaussian puff and urban cfd model for two
  large cities, Atmospheric Environment 39~(6) (2005) 1049--1068.

\bibitem{berchet2017cost}
A.~Berchet, K.~Zink, C.~Muller, D.~Oettl, J.~Brunner, L.~Emmenegger,
  D.~Brunner, A cost-effective method for simulating city-wide air flow and
  pollutant dispersion at building resolving scale, Atmospheric environment 158
  (2017) 181--196.

\bibitem{berchet2017evaluation}
A.~Berchet, K.~Zink, D.~Oettl, J.~Brunner, L.~Emmenegger, D.~Brunner,
  Evaluation of high-resolution gramm--gral (v15. 12/v14. 8) no x simulations
  over the city of z{\"u}rich, switzerland, Geoscientific Model Development
  10~(9) (2017) 3441--3459.

\bibitem{almbauer2000simulation}
R.~Almbauer, D.~{\"O}ttl, M.~Bacher, P.~Sturm, Simulation of the air quality
  during a field study for the city of graz, Atmospheric environment 34~(27)
  (2000) 4581--4594.

\bibitem{almbauer2000analysis}
R.~Almbauer, M.~Piringer, K.~Baumann, D.~Oettl, P.~Sturm, Analysis of the daily
  variations of wintertime air pollution concentrations in the city of graz,
  austria, Environmental Monitoring and Assessment 65~(1-2) (2000) 79--87.

\bibitem{fabbi2019impact}
S.~Fabbi, S.~Asaro, A.~Bigi, S.~Teggi, G.~Ghermandi, Impact of vehicular
  emissions in an urban area of the po valley by microscale simulation with the
  gral dispersion model, in: IOP Conference Series: Earth and Environmental
  Science, Vol. 296, IOP Publishing, 2019, p. 012006.

\bibitem{kurz2014projection}
C.~Kurz, R.~Orthofer, P.~Sturm, A.~Kaiser, U.~Uhrner, R.~Reifeltshammer,
  M.~Rexeis, Projection of the air quality in vienna between 2005 and 2020 for
  no2 and pm10, Urban climate 10 (2014) 703--719.

\bibitem{oettl2003dispersion}
D.~Oettl, P.~Sturm, R.~Almbauer, S.~Okamoto, K.~Horiuchi, Dispersion from road
  tunnel portals: comparison of two different modelling approaches, Atmospheric
  Environment 37~(37) (2003) 5165--5175.

\bibitem{oettl2002simple}
D.~Oettl, P.~J. Sturm, M.~Bacher, G.~Pretterhofer, R.~A. Almbauer, A simple
  model for the dispersion of pollutants from a road tunnel portal, Atmospheric
  Environment 36~(18) (2002) 2943--2953.

\bibitem{sturmpollutant}
P.~Sturm, D.~Oettl, M.~Bacher, R.~Almbauer, Pollutant dispersion in the
  vicinity of tunnel portals.

\bibitem{wolkinger2018evaluating}
B.~Wolkinger, W.~Haas, G.~Bachner, U.~Weisz, K.~W. Steininger, H.-P. Hutter,
  J.~Delcour, R.~Griebler, B.~Mittelbach, P.~Maier, et~al., Evaluating health
  co-benefits of climate change mitigation in urban mobility, International
  journal of environmental research and public health 15~(5) (2018) 880.

\bibitem{landesregierung2017documentation}
A.~D.~S. LANDESREGIERUNG, Documentation of the prognostic mesoscale model gramm
  (graz mesoscale model) vs. 17.1.

\bibitem{landesregierung2018documentation}
A.~D.~S. LANDESREGIERUNG, Documentation of the lagrangian particle model gral
  (graz lagrangian model) vs. 18.1.

\bibitem{optimization2017}
D.~Roddis, F.~Manansala, P.~Boulter, J.~Barnett, C.~Kurz,
  \href{https://www.chiefscientist.nsw.gov.au/__data/assets/pdf_file/0011/125030/ACTAQ-GRAL-optimisation-MAIN.pdf}{Optimisation
  of the application of gral in the australian context} (October 2017).
\newline\urlprefix\url{https://www.chiefscientist.nsw.gov.au/__data/assets/pdf_file/0011/125030/ACTAQ-GRAL-optimisation-MAIN.pdf}

\bibitem{optimization2017app}
D.~Roddis, F.~Manansala, P.~Boulter, J.~Barnett, C.~Kurz,
  \href{https://www.chiefscientist.nsw.gov.au/__data/assets/pdf_file/0019/125038/171101_ACTAQ-GRAL-optimisation-APPENDICES-A-F.pdf}{Optimisation
  of the application of gral in the australian context, appendices a-f}
  (October 2017).
\newline\urlprefix\url{https://www.chiefscientist.nsw.gov.au/__data/assets/pdf_file/0019/125038/171101_ACTAQ-GRAL-optimisation-APPENDICES-A-F.pdf}

\bibitem{massoud2011intraurban}
R.~Massoud, A.~L. Shihadeh, M.~Roumi{\'e}, M.~Youness, J.~Gerard, N.~Saliba,
  R.~Zaarour, M.~Abboud, W.~Farah, N.~A. Saliba, Intraurban variability of pm10
  and pm2. 5 in an eastern mediterranean city, Atmospheric Research 101~(4)
  (2011) 893--901.

\bibitem{saliba2010origin}
N.~Saliba, F.~El~Jam, G.~El~Tayar, W.~Obeid, M.~Roumie, Origin and variability
  of particulate matter (pm10 and pm2. 5) mass concentrations over an eastern
  mediterranean city, Atmospheric Research 97~(1-2) (2010) 106--114.

\bibitem{farah2018analysis}
W.~Farah, M.~M. Nakhl{\'e}, M.~Abboud, N.~Ziade, I.~Annesi-Maesano, R.~Zaarour,
  N.~Saliba, G.~Germanos, N.~A. Saliba, A.~L. Shihadeh, et~al., Analysis of the
  continuous measurements of pm 10 and pm 2.5 concentrations in beirut,
  lebanon., Environmental Engineering \& Management Journal (EEMJ) 17~(7).

\bibitem{lovett2018oxidative}
C.~Lovett, M.~H. Sowlat, N.~A. Saliba, A.~L. Shihadeh, C.~Sioutas, Oxidative
  potential of ambient particulate matter in beirut during saharan and arabian
  dust events, Atmospheric environment 188 (2018) 34--42.

\bibitem{daher2013chemical}
N.~Daher, N.~A. Saliba, A.~L. Shihadeh, M.~Jaafar, R.~Baalbaki, C.~Sioutas,
  Chemical composition of size-resolved particulate matter at near-freeway and
  urban background sites in the greater beirut area, Atmospheric environment 80
  (2013) 96--106.

\bibitem{mg2020novel}
M.~Ghadban, A.~Baayoun, I.~Lakkis, S.~Najem, N.~A. Saliba, A.~Shihadeh, A novel
  method to improve temperature forecast in data-scarce urban environments with
  application to the urban heat island in beirut, to appear in Urban Climate
  (2020).

\bibitem{baayoun2019emission}
A.~Baayoun, W.~Itani, J.~El~Helou, L.~Halabi, S.~Medlej, M.~El~Malki,
  A.~Moukhadder, L.~K. Aboujaoude, V.~Kabakian, H.~Mounajed, et~al., Emission
  inventory of key sources of air pollution in lebanon, Atmospheric Environment
  215 (2019) 116871.

\bibitem{alam2017tracking}
M.~Alam, J.~Powell, Tracking sustainable mobility.

\bibitem{jassimenvironmental}
D.~H.~M. Jassim, F.~H. Ibraheem, H.~A. Jasim, Environmental impact of
  electrical power generators in iraq (2016).

\bibitem{newsdeeply2014}
\href{https://www.newsdeeply.com/syria/articles/2014/04/02/aid-groups-say-government-opposition-both-blocking-aid-to-syrian-civilians}{News
  deeply, facing electricity cuts, {Aleppo Creates a Generator Economy}}
  (2014).
\newline\urlprefix\url{https://www.newsdeeply.com/syria/articles/2014/04/02/aid-groups-say-government-opposition-both-blocking-aid-to-syrian-civilians}

\bibitem{oguntoke2017degradation}
O.~Oguntoke, A.~Adeyemi, Degradation of urban environment and human health by
  emissions from fossil-fuel combusting electricity generators in abeokuta
  metropolis, nigeria, Indoor and Built Environment 26~(4) (2017) 538--550.

\bibitem{badaro2014geostatistical}
N.~Badaro-Saliba, J.~Adjizian-G{\'e}rard, R.~Zaarour, M.~Abboud, W.~Farah,
  A.~Saliba, A.~Shihadeh, A geostatistical approach for assessing population
  exposure to no 2 in a complex urban area (beirut, lebanon), Stochastic
  environmental research and risk assessment 28~(3) (2014) 467--474.

\bibitem{gpsv}
\href{http://www.gpsvisualizer.com}{Gps visualizer}.
\newline\urlprefix\url{http://www.gpsvisualizer.com}

\bibitem{european2016}
\href{https://www.eea.europa.eu/publications/emep-eea-guidebook-2016/part-b-sectoral-guidance-chapters/1-energy/1-a-combustion/1-a-4-small-combustion-2016/view}{European
  environment agency, emep/eea air pollutant emission inventory guidebook-small
  combustion} (2016).
\newline\urlprefix\url{https://www.eea.europa.eu/publications/emep-eea-guidebook-2016/part-b-sectoral-guidance-chapters/1-energy/1-a-combustion/1-a-4-small-combustion-2016/view}

\bibitem{wf}
\href{https://www.windfinder.com/windstatistics/beirut}{Wind finder}.
\newline\urlprefix\url{https://www.windfinder.com/windstatistics/beirut}

\bibitem{tradingeconomics}
\href{https://tradingeconomics.com/lebanon/electric-power-consumption-kwh-per-capita-wb-data.html}{Trading
  economics lebanon- electric power consumption (kwh per capita)} (2014).
\newline\urlprefix\url{https://tradingeconomics.com/lebanon/electric-power-consumption-kwh-per-capita-wb-data.html}

\bibitem{oke1988street}
T.~R. Oke, Street design and urban canopy layer climate, Energy and buildings
  11~(1-3) (1988) 103--113.

\bibitem{afiq2012effects}
W.~Afiq, C.~N. Azwadi, K.~Saqr, Effects of buildings aspect ratio, wind speed
  and wind direction on flow structure and pollutant dispersion in symmetric
  street canyons: a review, Int J Mech Mater Eng 7~(2) (2012) 158--165.

\bibitem{yazid2014review}
A.~W.~M. Yazid, N.~A.~C. Sidik, S.~M. Salim, K.~M. Saqr, A review on the flow
  structure and pollutant dispersion in urban street canyons for urban planning
  strategies, Simulation 90~(8) (2014) 892--916.

\end{thebibliography}

\newpage

\section{Appendices and Supplementary Material}\label{SecApp}
In this section, we present some additional information regarding the buildings and generators densities, the statistical work done on the generators' capacities, operation and spatial distribution and the vertical population density distribution along the floors of Beirut's buildings.

\subsection{Appendix A: Generators' Distribution}\label{appA}
\subsubsection*{First Order Approximation Generators' Distribution (GD2)}\label{appAGD1}
\begin{figure}[!htb]
	\begin{center}
		\includegraphics[width=\columnwidth]{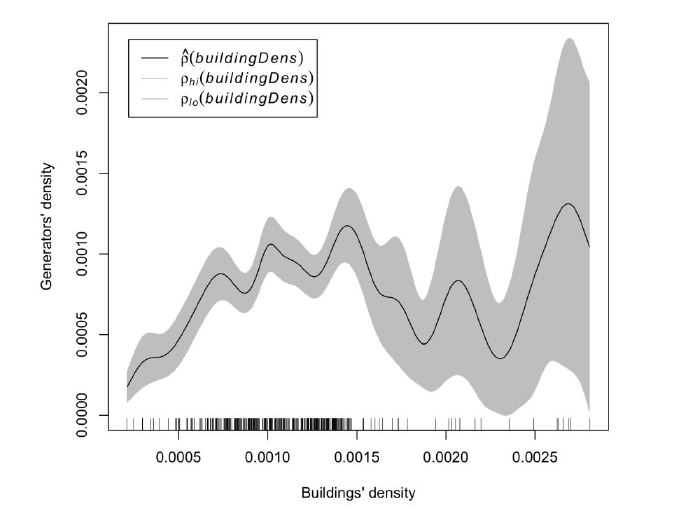}
		\caption{The density of generators as function of that of the buildings in m$^2$ with upper and lower bounds given by $\rho$$_{hi}$ and $\rho$$_{low}$}
		\label{buildensity}
	\end{center}
\end{figure}

The purpose of the survey is to infer the distribution of generators in the city beyond the study area. Their density is linked to a number of factors including buildings' density, their functions, the EDL zonings, population density, and socioeconomic conditions among others. 
The unavailability of a significant amount of the aforementioned data for Beirut restricted our analysis to exploring the generator's coordinates' dependence on a single covariate, which is buildings' locations and therefore limited the conclusions of the below presented analysis. However, we opted to present it for the sake of completeness. For this purpose, we treated the generators' coordinates as a spatial point process whose density, $\rho$$_{gen}$ in relation to that of the buildings, $\rho$$_{buil}$, is to be investigated. More precisely, we extracted the buildings' centroids and carried out the Sheather and Jones' kernel smoothing to get $\rho$$_{buil}$ in the study  area. A nonparametric smoothing estimate of $\rho$$_{gen}$ is fitted against $\rho$$_{buil}$, which is now a continuous spatial covariate. The result is given in Figure \ref{buildensity}.

\begin{figure}[!htb]
	\begin{center}
		\includegraphics[width=\columnwidth]{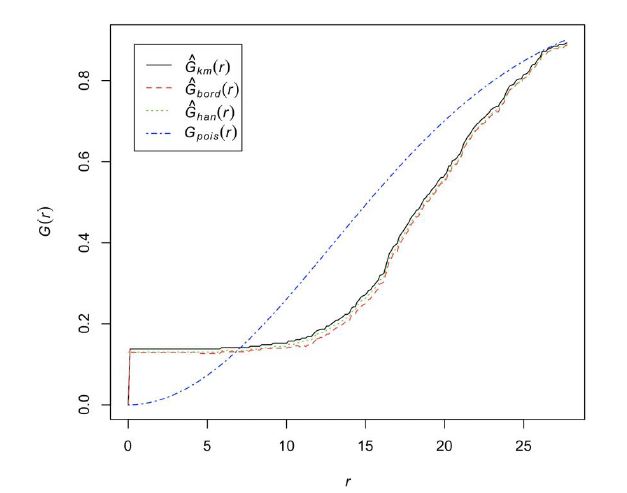}
		\caption{G(r) of the buildings}
		\label{grb}
	\end{center}
\end{figure}

\begin{figure}[!htb]
	\begin{center}
		\includegraphics[width=\columnwidth]{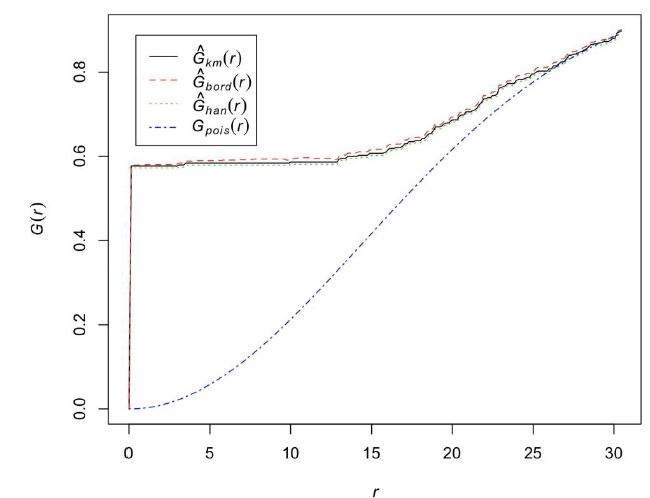}
		\caption{G(r) of the generators}
		\label{grg}
	\end{center}
\end{figure}

Relying on the above result exclusively would allow us to generate a number of generators corresponding to the buildings' density anywhere in the city. However, our study area is nearly socio-economically homogenous, which is considered an affluent neighborhood, with mixed and residential buildings, and with low population density compared to other parts of the city. Therefore, extrapolating $\rho_{gen}$, based on the above, undermines the effect of other covariates. More explicitly, the population density in our study area is higher than that in down-town Beirut for example, but on the other hand the latter has low buildings' density, near 0.005. Corniche mazraa has a high building density, near 0.0025 but is less affluent than our study area. \\
Additional clustering tests on the distribution of generators and that of the buildings also exhibited mismatching patterns, particularly the nearest neighbors distance function G as a function of distance r, (plotted against distance in Figures \ref{grb} and \ref{grg} for the buildings and generators respectively) which quantifies the deviation from complete spatial randomness. In the regions where the curve lies above the blue curve, which corresponds to a Poisson process, the points are spatially clustered whereas when the curve is below it the points are dispersed.\\
r

Given these spatial variabilities in covariates which we could not account for in the spatial analysis 587 we carried out, the latter becomes inconclusive. However, we decided to go for the ratio of 1:2 for the city as first order approximation, which is simply the ratio of the total number of generators to that of the buildings in the study area, which could be refined as data on other covariates is acquired.

\subsubsection*{Statistical Based Approximation Generators' Distribution (GD1) }\label{appAGD1}

\begin{figure}[!htb]
	\begin{center}
		\includegraphics[width=\columnwidth]{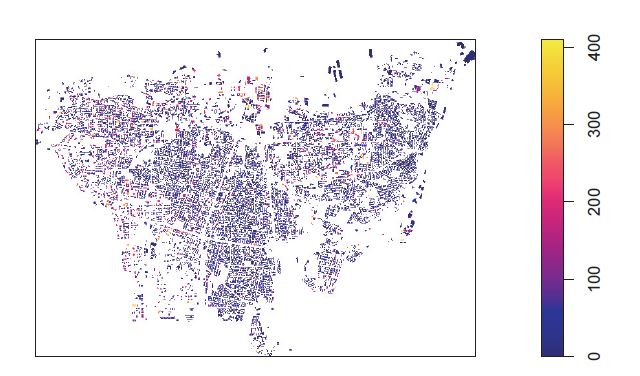}
		\caption{The population raster generated using the EDL consumptions, buildings' heights, and average consumption per capita.}
		\label{popraster}
	\end{center}
\end{figure}

We first transformed the locations of the generators into an unmarked point process:this allowed us to model the density of the generators $\rho$ as a function of the population N(x; y), which is our single covariate inferred from the electricity consumption bill and the buildings' heights as shown in Figure \ref{popraster}.

\begin{figure}[!htb]
	\begin{center}
		\includegraphics[width=\columnwidth]{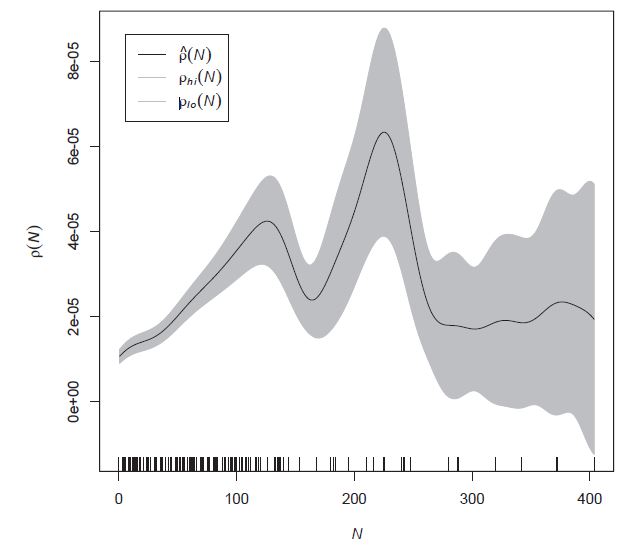}
		\caption{The density of generators per meter squared versus the population in the surveyed area.}
		\label{poprho}
	\end{center}
\end{figure}

\begin{figure}[!htb]
	\begin{center}
		\includegraphics[width=\columnwidth]{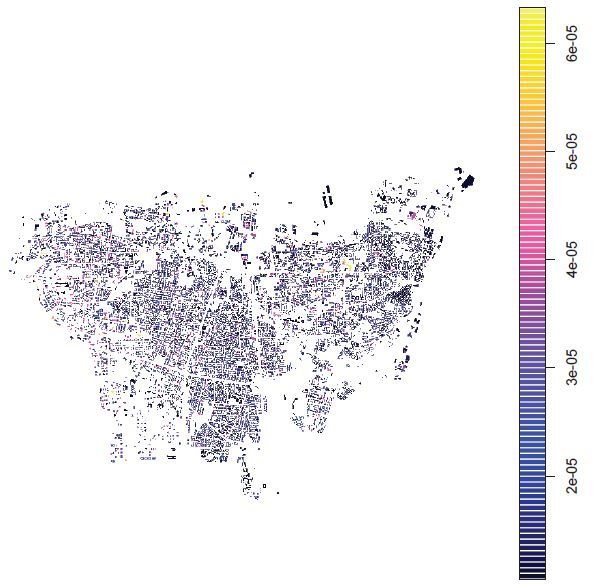}
		\caption{The predicted density of generators per meter squared versus the population in the city.}
		\label{poprhop}
	\end{center}
\end{figure}

For this first part we just cared about their spatial distribution neglecting their generation capacities. The latter was modeled as a multi-type marked point process, which will be explained in the coming section. We sought a non-parametric relation between $\rho$ and N(x; y), which was used to predict the density of generators beyond the surveyed area. $\rho$ is given in Figure \ref{poprho} and its prediction is given in Figure \ref{poprhop}.

\begin{figure}[!htb]
	\begin{center}
		\includegraphics[width=\columnwidth]{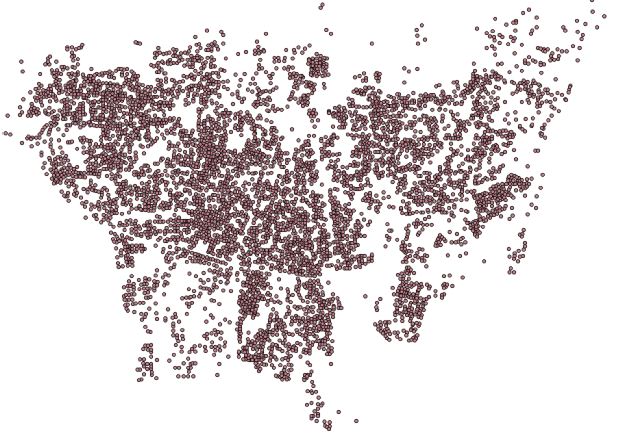}
		\caption{The predicted locations with fractional numbers.}
		\label{loc}
	\end{center}
\end{figure}

This raster then was transformed into points by multiplying $\rho$ assigned to each building with its corresponding area as show in Figure \ref{loc}. Note that this gives fractions, so a location can have 0.2 generators. The number of generators with values greater than a given $\rho$ is shown in the Table \ref{Tablesara}. The idea is to lump nearby locations with fractional numbers and replace them with their equivalent.

\begin{table}[!ht]
	\centering
	\begin{tabular}{cccccc}
		\toprule
		Fraction	& $>0.01$	& $>0.05$	& $>0.1$	& $>0.5$ & $>1$ \\
		n	& 6052	&  1540	& 619 & 	74& 27 \\
		\bottomrule
	\end{tabular}
	\caption{Number of generators versus a threshold in the fractions.}
	\label{Tablesara}
\end{table}

\begin{figure}[!htb]
	\begin{center}
		\includegraphics[width=\columnwidth]{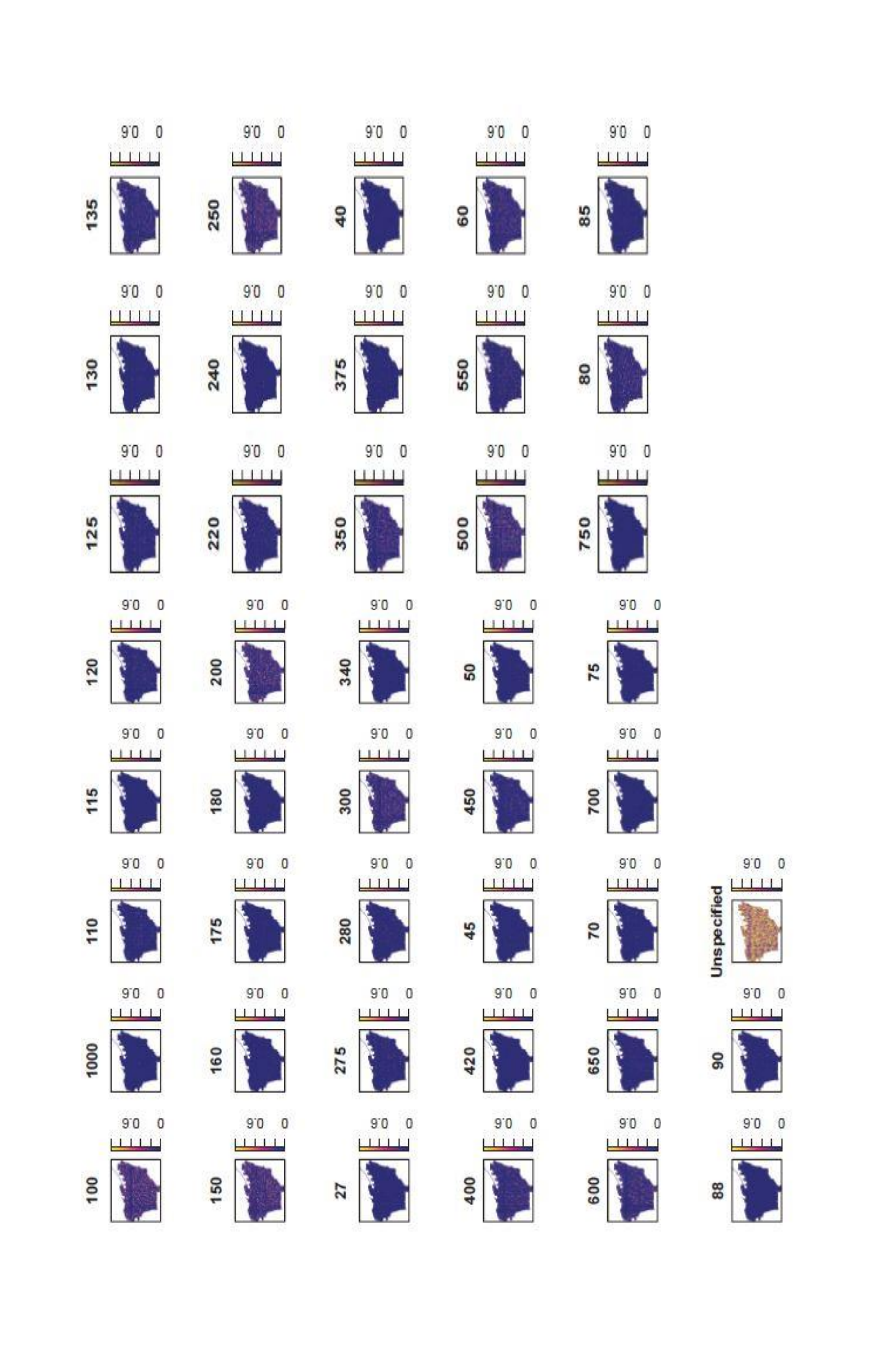}
		\caption{The predicted probability of the presence of a generator at a given location for all capacities.}
		\label{prob}
	\end{center}
\end{figure}

\begin{figure}[!htb]
	\begin{center}
		\includegraphics[width=\columnwidth]{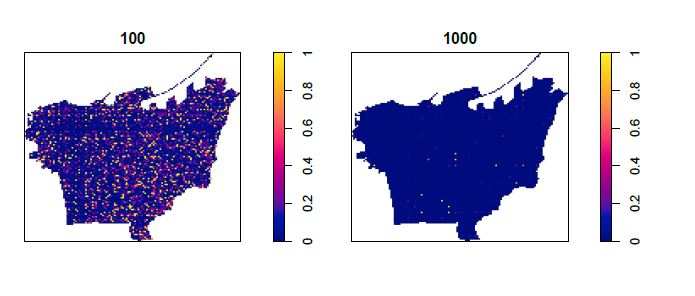}
		\caption{A sample of the predicted probabilities of the presence of a generators for a 100 and 1000 kVA.}
		\label{predprob}
	\end{center}
\end{figure}

In order to assign a capacity to the predicted locations we looked at the points as a marked process. Having the marks in the surveyed area we could investigate their interdependence and conditional probabilities: that is the probability of finding a generator with
capacity ci at location ri given capacities cj and the locations rj of the others. This gave us the relative probabilities below shown in Figure \ref{prob} for all the capacities and a sample for 100 and 1000 kVA in Figure \ref{predprob}.

\begin{figure}[!htb]
	\begin{center}
		\includegraphics[width=\columnwidth]{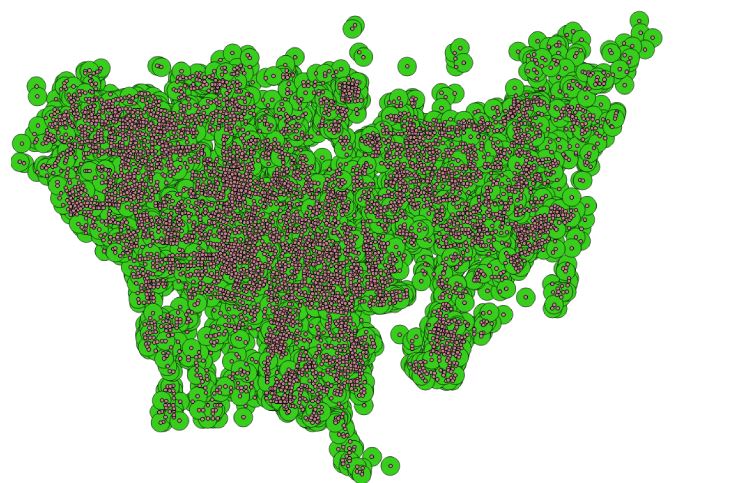}
		\caption{Buffer zones around predicted fractional numbers.}
		\label{buffer}
	\end{center}
\end{figure}

We should note that these probabilities do not depend on the population covariate. They are just the resultant of analysis of the interaction" between the surveyed generators' locations. The idea is to add all the rasters from Figure \ref{prob} and overlay the result on the density raster from Figure \ref{poprhop}, which was inferred using the population after transforming the fractions into real numbers using the buffer zones as shown in Figure \ref{buffer}.

\subsection{Appendix B: Population Density Distribution}\label{appC}
\begin{figure}[!htb]
	\begin{center}
		\includegraphics[width=\columnwidth]{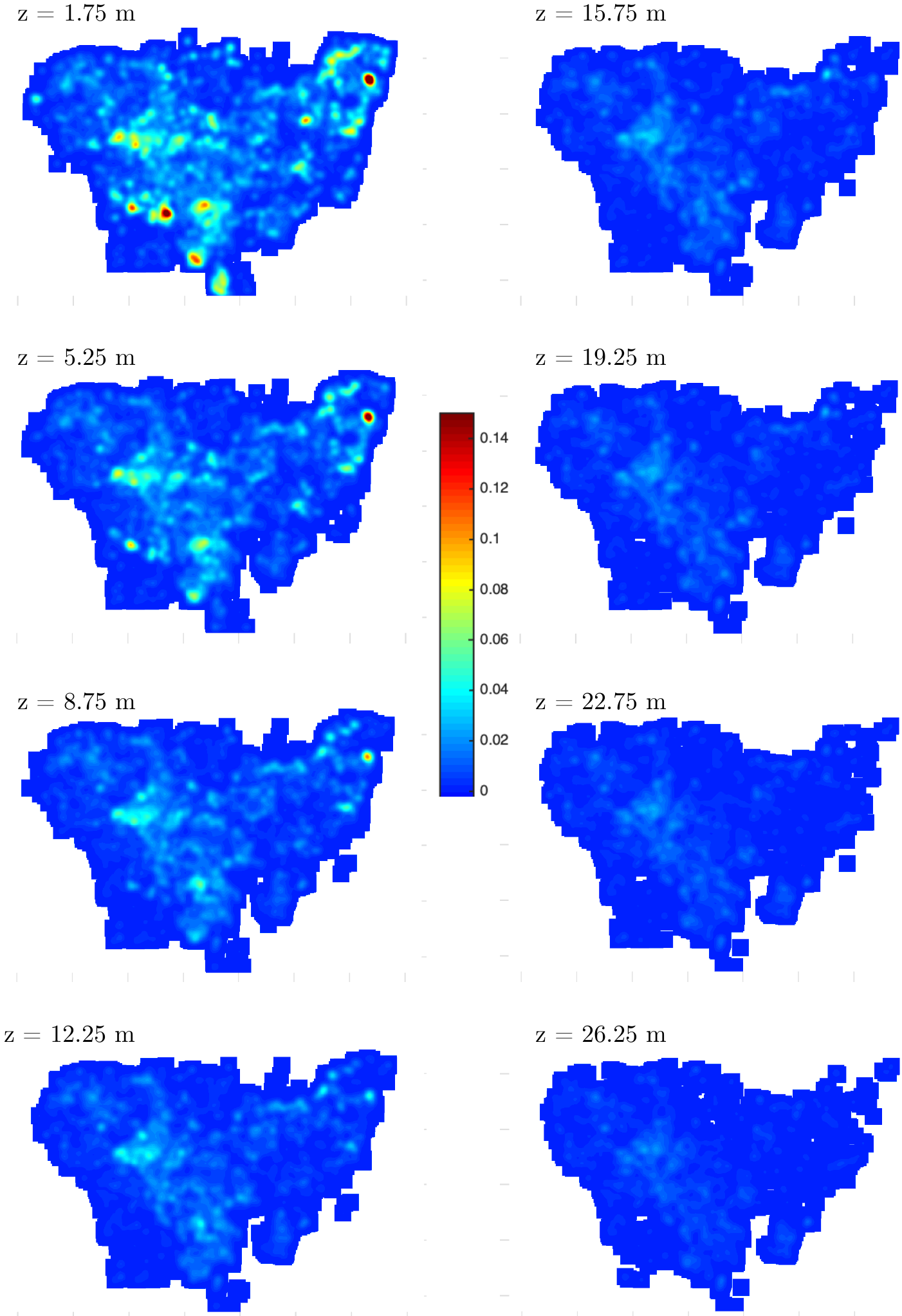}
		\caption{Population density distribution for the first eight floors over the (x,y) plane of Beirut. The results are artificially smoothed for clarity.}
		\label{smoothedpop}
	\end{center}
\end{figure}

The population density distribution in each building in Beirut was obtained from the data provided by EDL. For each building, the number of the floors is deduced based on its height and an average floor height of 3.5 meters. Then, the population is distributed uniformly along the floors of each building. The obtained results are artificially smoothed and represented over the (x,y) plane of Beirut for the first eight floors in Figure \ref{smoothedpop}. The z values represented in this figure refer to the elevation of the center of each floor.

\end{document}